\documentclass[twocolumn]{aastex631}
\usepackage{amsmath, amsthm, amssymb, amsfonts}
\usepackage{enumitem,kantlipsum}

\emergencystretch=\maxdimen
\usepackage{graphicx}

\begin{document}
\title{The \textit{JWST} Early Release Science Program for Direct Observations of Exoplanetary Systems IV: \\ NIRISS Aperture Masking Interferometry Performance and Lessons Learned}

\author[0000-0001-6871-6775]{Steph Sallum}
\altaffiliation{Send manuscript correspondence to S.S., ssallum@uci.edu}
\affiliation{Department of Physics and Astronomy, University of California, Irvine, 4129 Frederick Reines Hall, Irvine, CA, USA}

\author[0000-0003-2259-3911]{Shrishmoy Ray}
\affiliation{University of Exeter, Astrophysics Group, Physics Building, Stocker Road, Exeter, EX4 4QL, UK}

\author[0000-0003-2769-0438]{Jens Kammerer}
\affiliation{European Southern Observatory, Karl-Schwarzschild-Straße 2, 85748 Garching, Germany}

\author{Anand Sivaramakrishnan}
\affiliation{Space Telescope Science Institute, 3700 San Martin Drive, Baltimore, MD 21218, USA}

\author{Rachel Cooper}
\affiliation{Space Telescope Science Institute, 3700 San Martin Drive, Baltimore, MD 21218, USA}

\author[0000-0002-7162-8036]{Alexandra Z.~Greebaum}
\affiliation{IPAC, California Institute of Technology, 1200 E. California Blvd., Pasadena, CA 91125, USA}

\author{Deepashri Thatte}
\affiliation{Space Telescope Science Institute, 3700 San Martin Drive, Baltimore, MD 21218, USA}

\author[0000-0003-1863-4960]{Matthew De Furio}
\affiliation{Department of Astronomy, The University of Texas at Austin, 2515 Speedway Stop C1400, Austin, TX 78712, USA}

\author[0000-0002-8332-8516]{Samuel M. Factor}
\affiliation{Department of Astronomy, The University of Texas at Austin, 2515 Speedway Stop C1400, Austin, TX 78712, USA}

\author[0000-0003-1227-3084]{Michael R.~Meyer   }\affiliation{Department of Astronomy, University of Michigan, 1085 S. University, Ann Arbor, MI 48109, USA}

% ...

\begin{abstract}
We present a performance analysis for the aperture masking interferometry (AMI) mode on board the \textit{James Webb Space Telescope} Near Infrared Imager and Slitless Spectrograph (\textit{JWST}/NIRISS).
Thanks to self-calibrating observables, AMI accesses inner working angles down to and even within the classical diffraction limit. 
The scientific potential of this mode has recently been demonstrated by the Early Release Science (ERS) 1386 program with a deep search for close-in companions in the HIP 65426 exoplanetary system. 
As part of ERS 1386, we use the same dataset to explore the random, static, and calibration errors of NIRISS AMI observables.
We compare the observed noise properties and achievable contrast to theoretical predictions. 
We explore possible sources of calibration errors, and show that differences in charge migration between the observations of HIP 65426 and point-spread function calibration stars can account for the achieved contrast curves. 
Lastly, we use self-calibration tests to demonstrate that with adequate calibration, NIRISS F380M AMI can reach contrast levels of $\sim9-10$ mag at $\gtrsim\lambda/D$.
These tests lead us to observation planning recommendations and strongly motivate future studies aimed at producing sophisticated calibration strategies taking these systematic effects into account. 
This will unlock the unprecedented capabilities of \textit{JWST}/NIRISS AMI, with sensitivity to significantly colder, lower mass exoplanets than lower-contrast ground-based AMI setups, at orbital separations inaccessible to \textit{JWST} coronagraphy.

\end{abstract}

\section{Introduction}\label{sec:intro}
The \textit{James Webb Space Telescope} \citep[\textit{JWST};][]{2006SSRv..123..485G,2023PASP..135f8001G} is now the first space-based infrared interferometer thanks to the aperture masking interferometry (AMI) mode on the Near Infrared Imager and Slitless Spectrograph (NIRISS) instrument \citep[e.g.][]{2012SPIE.8442E..2SS,2022SPIE12180E..3NK,2023PASP..135a5003S}. 
AMI transforms a conventional telescope into an interferometric array via a pupil-plane mask \citep[e.g.][]{2000PASP..112..555T}.
By blocking the majority of the light, AMI enables the calculation of self-calibrating observables that provide sensitivity to asymmetries close to, and even within the classical diffraction limit. 
This technique has been thoroughly demonstrated from the ground via observations of stellar and substellar companions \citep[e.g.][]{2008ApJ...678L..59I,2012ApJ...753L..38B,2015ApJ...806L...9H}, circumstellar disks \citep[e.g.][]{2001Natur.409.1012T,2019ApJ...883..100S,2023ApJ...953...55S} and more. 

Theoretical predictions of \textit{JWST}/NIRISS AMI contrast have highlighted its potential for direct exoplanet studies \citep[e.g.][]{2020SPIE11446E..11S}, including observations of actively forming planets \citep[e.g.][]{2019JATIS...5a8001S} and young planets near the water ice line that are colder than those accessible with ground-based AMI \citep[e.g.][]{2023MNRAS.519.2718R}. 
\textit{JWST}/NIRISS AMI has now been explored as part of the Early Release Science (ERS) 1386 program \citep[described in its entirety in][]{2022PASP..134i5003H}.
In \citet{ray_subm}, we demonstrate the unique science capabilities of  NIRISS AMI, with a deep non-detection that places limits on close-in planets in the HIP 65426 exoplanetary system.
Here we present an accompanying study to assess in detail the achievable contrast and noise characteristics of the NIRISS AMI mode. 

\subsection{NIRISS Aperture Masking Interferometry}
On NIRISS, the AMI mode passes light from seven of the 18 \textit{JWST} primary mirror segments.
The images on the detector are then the sums of interference fringes from the resulting 21 hole pairs (baselines), each of which probes a unique spatial frequency.
Analyzing these data via Fourier techniques or fringe fitting enables the calculation of complex visibilities, which are the amplitudes and phases of the fringes associated with the mask baselines. 
From the complex visibilities we can calculate closure phases \citep[e.g.][]{1986Natur.320..595B}, sums of phases around baselines that form triangles, and squared visibilities, the powers associated with the various baselines \citep[e.g.][]{1958MNRAS.118..276J}. 
These and related quantities can be used to infer the source brightness distribution via model fitting and image reconstruction \citep[often applied together to understand the limitations of each approach][]{2017ApJS..233....9S}.

\subsection{Outline of this Paper}
Here we discuss in detail the data reduction steps taken in the ERS 1386 AMI observations, their achieved contrast, and possible limiting noise sources for NIRISS AMI based on the analysis of the ERS 1386 data. 
The outline of the paper is as follows:
In Section \ref{sec:noise} we provide an overview of AMI noise sources, and introduce the theoretical framework that we use to quantify AMI noise and performance.
In Section \ref{sec:obs} we briefly describe the observations of HIP 65426 and its two PSF references. 
In Section \ref{sec:red_an} we describe and justify the AMI data reduction steps that were taken. 
There we also describe how we generate the contrast curves presented here and in \citet{ray_subm}.
In Section \ref{sec:perf} we discuss the observed performance of the ERS 1386 AMI observations, and use a variety of tests to explore the limiting noise sources for NIRISS AMI. 
We discuss the results of these tests more generally in Section \ref{sec:disc} before concluding with recommendations for future NIRISS AMI programs in Section \ref{sec:conc}.

\section{AMI Noise Sources}\label{sec:noise}
AMI's sensitivity to close-in asymmetries is largely enabled by self-calibrating phase observables \citep[e.g. closure phase;][]{1986Natur.320..595B}. 
Due to the importance of phase and closure phase for achieving AMI resolution, in this work we focus primarily on phase and closure phase noise properties, leaving amplitude characterization to future studies. 
Here we briefly describe different sources of phase errors in AMI data, and establish the theoretical framework that we use to assess and predict NIRISS AMI performance. 
Following \citet{2013MNRAS.433.1718I},below  we describe three types of AMI phase noise sources: random, static, and calibration errors.

\subsection{Random Phase Errors}\label{sec:rand}
Random errors are noise sources that contribute to variations around the mean.
These include shot noise from the star and background, and detector noise sources such as read noise and dark current. 
For an $N_h$-hole mask, we express the complex visibility phase scatter induced by these random noise terms in a manner consistent with \citet{2013MNRAS.433.1718I}:
\begin{equation}
    \sigma(\phi) = \frac{N_h}{N_p V} \sqrt{0.5(N_p + n_p\sigma_{pix})}. \label{eq-pn}
\end{equation}
where $V$ is the fringe visibility of the observation compared to a perfect point source, and $N_p$ is the total number of photons collected from the science target.

The $\sigma_{pix}$ and $n_p$ terms capture detector-level noise contributions, with $\sigma_{pix}$ giving the pixel-level noise in units of $\mathrm{e^-}$.
We use a general term for $\sigma_{pix}$ since measurements of NIRISS' combined read noise, 1/f noise \citep[correlated noise between pixels caused by the readout electronics configuration; e.g.][]{2020AJ....160..231S}, and dark current noise exist in units of $\mathrm{e^-}$.\footnote{https://jwst-docs.stsci.edu/jwst-near-infrared-imager-and-slitless-spectrograph/niriss-instrumentation/niriss-detector-overview/niriss-detector-performance} 
We note that the magnitude of these combined detector noise terms is low enough that, for the observations conducted here, the $N_p$ term will dominate the expression for $\sigma(\phi)$.
We include the $\sigma_{pix}$ term for completeness and since this noise source has been well characterized for NIRISS. 

Lastly, the $n_p$ term is the total number of pixels in the image(s).
For example, if two $80\times80$ NIRISS AMI subframes were used to collect a total of $N_p$ photons, $n_p$ would be set to $2\times80\times80$. 
Furthermore, any windowing (e.g. with a super-Gaussian; see Section \ref{sec:red_an}) effectively decreases the number of pixels $n_p$, since windows suppress detector level noise where their throughput is low. 
When we apply a window function, we thus calculate $n_p$ by summing the pixels with the window throughput as a weight for each pixel.  
Extending Equation \ref{eq-pn} to closure phase, and allowing for each baseline to have its own visibility ($V_i$), the random closure phase noise for a single triangle is
\begin{equation}
    \sigma(\phi_{CP}) = \frac{N_h}{N_p}\sqrt{[0.5(N_p+n_p\sigma_{pix})]\sum_{i=1}^{3}\frac{1}{V_i^2}}.  \label{eq-cp}
\end{equation}

Here we are neglecting photon noise contributions from the thermal background, since this noise contribution is small and difficult to accurately measure given the small size of the NIRISS AMI subframes. 
We also note that random temporal phase variations and flat-field errors (if images are not acquired on identical pixels) can contribute to statistical closure phase errors. 
We neglect these in our theoretical statistical error expression since \textit{JWST's} stability should mean that these effects do not dominate the error budget. 
Furthermore, by following the recommended NIRISS AMI observing strategy of placing science targets and calibrators on identical pixels, flat-field errors can be mitigated.
We discuss this simplification in the context of the observed phase scatter in Section \ref{sec:randerrs}.

\subsection{Static Phase Errors}
Static errors cause the mean closure phase signal to deviate from zero.
Following the treatment in \citet{2013MNRAS.433.1718I}, we describe these errors as being third-order and higher in phase (after Taylor expanding the expression for the complex visibility phase).
These can arise from complex spatial variations in the wavefront phase across each mask subaperture. 
Time-invariable correlated and asymmetric detector systematics such as 1/f noise and alternating column pattern noise \citep[e.g.][]{2017PASP..129j5003R} can also contribute to static closure phase errors.
These higher-order static terms must be adequately calibrated to maximize AMI achievable contrast, which is why traditional AMI campaigns (including ERS 1386) observe dedicated point-spread function (PSF) reference targets.

\subsection{Calibration Errors}
If the static noise sources described above are not adequately removed during the calibration process, then the residual ``calibration error" will cause the achievable contrast to differ from the photon noise expectations defined in Equations \ref{eq-pn} and \ref{eq-cp}.
There are a variety of sources of calibration errors, including quasi-static speckles, which are slowly-varying spatial aberrations.
For \textit{JWST} such quasi-static errors could be caused by thermal drift of the telescope between observations of the science target and the PSF stars.
Calibration errors can also be caused by wavelength-dependent interactions between the light and optics that cause PSF shapes (and thus asymmetries) to depend on the target spectral type.
Any detector-level systematics that change from science target to calibrator would also induce calibration errors (with variability from frame to frame for a single object causing inflated random errors, e.g. Section \ref{sec:rand}).

\section{Observations}\label{sec:obs}
The ERS 1386 AMI observations were executed in the F380M filter on July 30, 2022 UTC (Table \ref{tab:obs}). 
The NISRAPID readout mode was used, and the images were acquired using the $80\times80$-pixel SUB80 subarray. 
Observations of HIP 65426 were preceded by observations of the PSF reference HD 115842 and followed by observations of a second PSF reference HD 116084. 
The HIP 65426, HD 115842, and HD 116084 integrations consisted of 13, 2, and 3 groups up the ramp, respectively. 
The numbers of total integrations recorded for the three objects were 10950, 15500, and 16000, respectively. 
As described in \citep{ray_subm}, the PSF references were chosen to be close on-sky ($<5^\circ$ separation) and with similar spectral types (B0.5-2I) to HIP 65426 (A2V).
They were executed in a non-interruptible sequence and placed on identical detector pixels as the science target to minimize calibration errors.

\begin{table}
\centering
\caption{Summary of Observations}\label{tab:obs}
\begin{tabular}{cccccccccc}
\hline
\hline

\multicolumn{1}{c}{\textbf{Star}} & \textbf{Readout} & \textbf{N\textsubscript{groups}}  & \textbf{N\textsubscript{ints}} & \textbf{t\textsubscript{exp} (s)}     \\

\hline
HD\,115842  & NISRAPID & 2 & 10000  & 2468.00    \\
                        & NISRAPID & 2 & 5500  & 1357.40  \\
HIP\,65426     & NISRAPID & 13 & 10000   &10766.40   \\
                        & NISRAPID & 13 & 950  &1022.81  \\
HD\,116084    & NISRAPID & 3 & 10000  &3222.4  \\
                        & NISRAPID & 3 & 6000  & 1933.44    \\
\hline
\end{tabular}
\end{table}

\section{NIRISS AMI Data Reduction and Analysis}\label{sec:red_an}
\subsection{Detector Level Corrections}\label{sec:det}
We use the \texttt{jwst} pipeline \citep[version 1.7.1;][]{bushouse_2022_7059013} to perform standard detector-level corrections for each group. These include superbias, reference pixel, linearity, persistence, and dark current corrections.
For any given reduction, we can apply some subset of the above corrections to the group level data products.
We then use stage 1 of \texttt{jwst} (i.e. by running \texttt{calwebb\_detector1}) to generate rate images, which serve as the starting point for custom AMI reduction steps using \texttt{SAMpy} \citep[e.g.][]{2022SPIE12183E..2MS}. 
We tested various combinations of the \texttt{jwst} detector-level corrections, and did not find significant differences between the AMI observables. 
The final reduction presented here and in \citep{ray_subm} thus includes all of the \texttt{jwst} reduction steps. 

\begin{figure}
    \centering
    \includegraphics[width=\columnwidth]{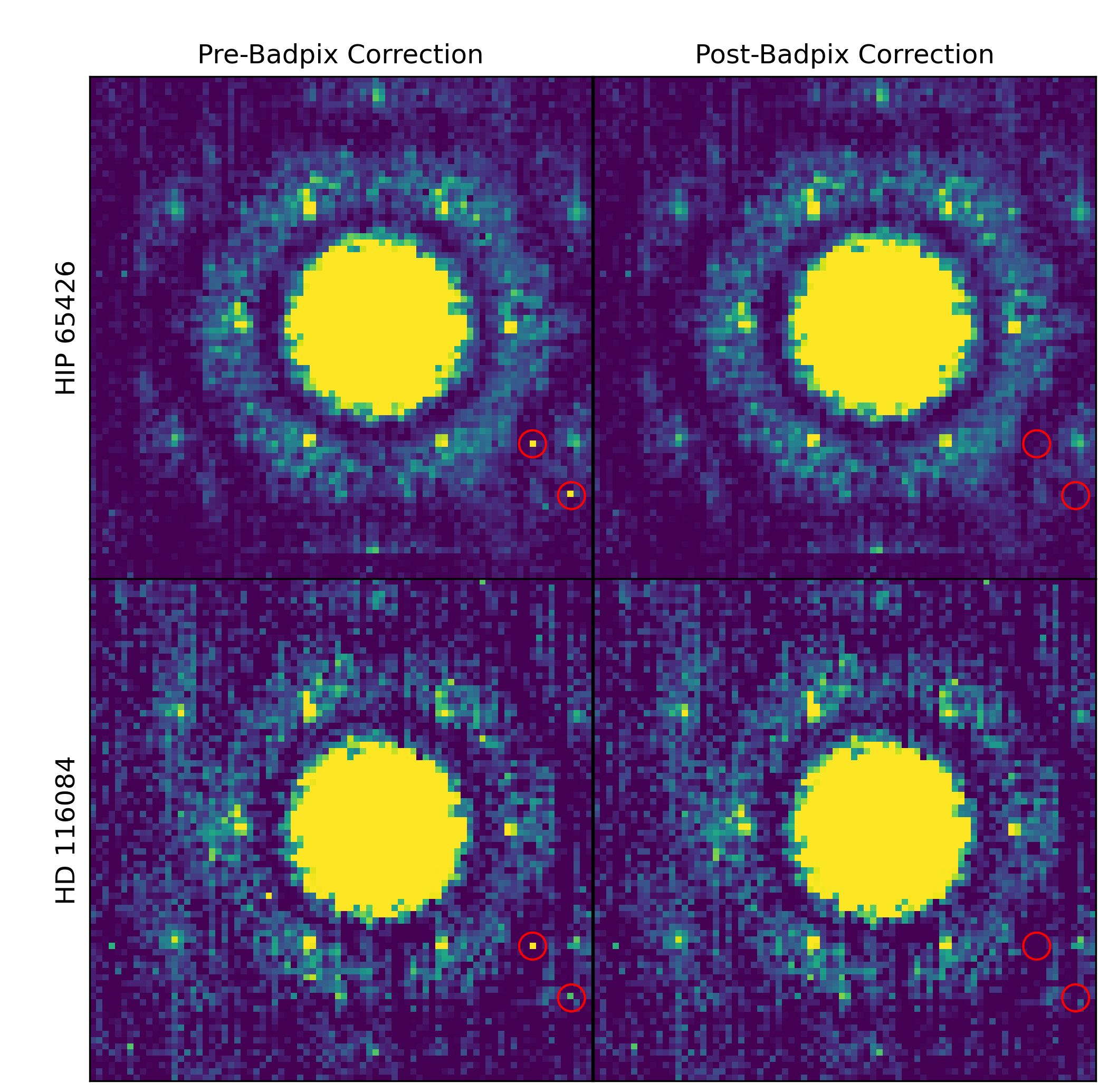}
    \caption{Example images of HIP 65426 (top panels) and HD 116084 (bottom panels) before and after bad pixel correction (left and right panels, respectively). Red circles indicate two bad pixels in the left panels that have been corrected in the right panels. The color scale has been stretched so that it ranges from 0 to 1$\%$ of the peak counts in each image. Residual striping features are seen in the HD 116084 images, which we discuss in Section \ref{sec:det}. The pixel size in all images is 65 mas, for a total field of view of 5.2 $\times$ 5.2 arcseconds.}
    \label{fig:badpix}
\end{figure}

We use the \texttt{jwst}-generated data quality, saturation, and jump detection arrays as the starting point for custom bad-pixel corrections.
We also flag additional bad pixels applying our own criteria:
\begin{enumerate}[leftmargin=*]
    \item Pixels in individual integrations that deviate by more than 5$\sigma$ from their median value (measured across all integrations). This selects high and low value outliers compared to an individual pixel's median behavior.
    \item Pixels in individual integrations that differ by $>100\times$ from their location in a median-filtered image with a 3x3 kernel size. This effectively selects high and low value outliers compared to pixel neighbors in individual images.
    \item Pixels with standard deviations (measured across all integrations) that differ by 5$\sigma$ compared to their neighboring pixels. This selects ``flaky" pixels by comparing a pixel's variability across all images to that of its immediate neighbors. 
\end{enumerate}
Together these criteria flag $1.5-2.2\%$ of pixels depending on the object.
We apply a bad pixel correction in Fourier space following the methodology described in \citet{2019MNRAS.486..639K} (i.e. minimizing the bad pixels' contributions to the Fourier-plane power outside the spatial frequencies sampled by the pupil).
Figure \ref{fig:badpix} shows example frames before and after bad pixel correction for HIP 65426 and the PSF reference HD 116084. 

Figure \ref{fig:badpix} also illustrates another low-level detector effect that can contribute to noise in the AMI observables. 
These are the sub-percent level vertical striping features seen in the PSF reference stars, which were observed with fewer groups than the science target. 
These may be caused by 1/f noise and/or alternating column pattern noise \citep[e.g. Section \ref{sec:noise};][]{2017PASP..129j5003R}. 
Such effects can be mitigated by the use of a window function that tapers the image away from the center of the PSF. 
We thus explore the use of a super-Gaussian window function, which has the following form: $\exp(-(r/\sigma)^n)$, where $r$ is the radius from the center of the image, and $\sigma$ determines the full width at half maximum (FWHM).
For all of the tests that follow and for the final reduction we use a radial exponent of $n=4$ for the super-Gaussian. 

\begin{figure*}
    \centering
    \includegraphics[width=0.8\textwidth]{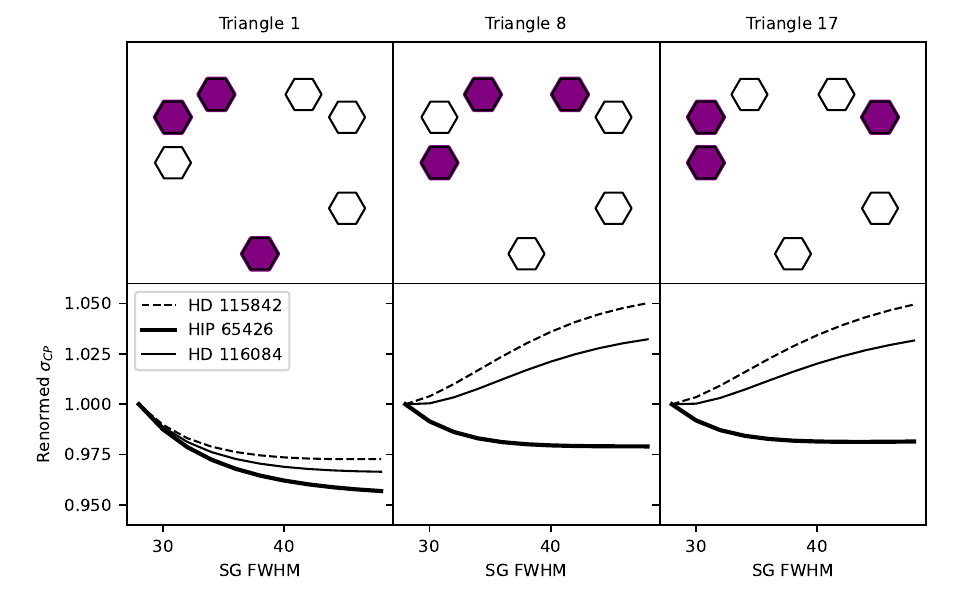}
    \caption{Top panels depict three example closing closing triangles from the NIRISS AMI mask. The black outlines show the positions of all seven holes, and the filled purple hexagons show the holes used to calculate the closure phases in the bottom panels. Bottom panels show the closure phase standard deviation around the mean ($\sigma_{CP}$) as a function of super-Gaussian FWHM (SG FWHM), for the closing triangles depicted in the top panels. The $\sigma_{CP}$ values are normalized by the value for the smallest SG FWHM of 28 pixels. The thick solid line shows HIP 65426. The thin solid and dashed lines respectively show the two PSF references HD 116084 and HD 115842. Increasing the super-Gaussian FWHM size for triangles with horizontal baselines leads to an increase in the random error of the PSF reference closure phases. This is because the larger super-Gaussian FWHMs allow for increased contributions by vertical striping on the detector (which has the largest effect on the horizontal baselines).}
    \label{fig:sgtest}
\end{figure*}

An aggressive super-Gaussian (e.g. small FWHM) would mean masking some portion of the PSF.
For reference, the extent of the central portion of the AMI interferogram (determined by the first null in the PSF of a single mask hole) has a size of $\sim36$ pixels in diameter.
The outer, ring-like portion of the interferogram, which corresponds to the first Airy ring in a single mask hole PSF (where signal is visible in each image) exists between $\sim36$ and $\sim66$ pixels in diameter. 
Thus, super-Gaussian functions with FWHMs significantly smaller than 36 pixels will attenuate the central portion of the interferogram, and those with FWHMs smaller than 66 pixels will attenuate the outer ring. 

Given the dimensions described above, the choice of super-Gaussian FWHM must be made to balance including more photons (large FWHM) against eliminating pixels that contribute excess noise (small FWHM).
We evaluate the choice of FWHM by quantifying the level of random error in the closure phases.
To do this, we measure the scatter around the mean of each closure phase by taking the standard deviation across all of the integrations for each object. 
We explore super-Gaussian FWHMs between 28 pixels and 48 pixels, and normalize the standard deviations by that for the smallest FWHM of 28 pixels. 
The upper FWHM limit of 48 pixels is chosen so that the attenuation at large distances is high enough to avoid systematics due to sharpness at the edge of the $64\times64$ pixel centered and cropped images.
A change in FWHM that increases the signal-to-noise ratio would lead to a decrease in the scatter around the mean, and vice versa. 

Figure \ref{fig:sgtest} shows examples of the results of these tests. 
For the majority of the closing triangles, the scatter about the mean decreases as the size of the super-Gaussian window increases. 
However, this is not the case for closure phases whose triangles include horizontal baselines, in the two PSF reference datasets.
This is because the spatial frequencies probed by horizontal baselines are particularly sensitive to asymmetries caused by vertical striping. 
The effect is not seen in the science target closure phases with horizontal baselines, because this striping is mitigated when more groups are used up the ramp.

The results shown in Figure \ref{fig:sgtest} seem to motivate applying different windowing settings to different closure phase calculations.
However, as we discuss in Section \ref{sec:fobs}, the noise in the calibrated closure phases was minimized in the reduction where all objects and closing triangles have the 48-pixel FWHM window applied. 
Furthermore, the possibility of inflated systematic errors due to this striping in the PSF references, and the resulting calibration errors, motivates detector-level removal of the stripes. 
We thus tested a simple destriping approach similar to the channel bias correction described in \citet{2017ApJS..233....9S}.
We used the outskirts of the images to estimate individual column biases which we then subtracted, but this did not lower the closure phase scatter for these triangles.
This is perhaps due to the fact that the SUB80 subarray is relatively small compared to the NIRISS AMI PSF. 
In future work more sophisticated destriping approaches will be explored.

\subsection{Extraction of Fourier Observables}\label{sec:ext}
\texttt{SAMpy} can extract Fourier observables by sampling either single pixels at the center of each baseline's location in Fourier space or multiple pixels for each baseline, with an extent defined by the user. 
The reductions presented here and in \citet{ray_subm} take the second approach, sampling all pixels within 25\% of the peak signal for each baseline. 
This flux cut is arbitrary and changeable; a 25\% value ensures that most of the (u,v) space for each baseline is sampled, but without beginning to sample nearby baselines.
Sampling the Fourier transform within a 50\% flux cut did not change the results significantly. 

The visibility amplitudes, visibility phases, and squared visibilities are calculated by averaging the observables measured from these regions of the Fourier transform. 
When using multiple pixels, the closure phases can be calculated in two ways: (1) by finding all individual triangles of pixels that close, calculating their bispectra, and averaging the bispectra for each closing triangle before taking the phase as the closure phase; or (2) by calculating an average complex visibility for each baseline in the closing triangle, calculating one bispectrum value using the three average visibilities, and then taking its phase as the closure phase. 

While closure phase approach (1) is strictly correct, it can be computationally expensive since many closing triangles of pixels may connect the three baselines. 
We thus test whether there is a significant difference between the observables calculated between approaches (1) and (2), and find that they are nearly identical. 
This may be because we first apply a window function that creates correlations between pixels in Fourier space.
We thus apply approach (2) for the closure phase tests shown here, and for the final reduction presented in \citet{ray_subm}.
However, we note that an additional advantage of approach (1) is that it can be applied without averaging all the pixel-triangle bispectra, to preserve phase information that changes within the mask subapertures.
This may be useful for trying to recover wide-separation signals that would be aliased by the minimum mask baselines, such as the signal from the wide-separation companion HIP 65426 b \citep[e.g.][]{ray_subm}. 

For all observables, we assign statistical error bars capturing the random phase and amplitude errors. 
We estimate these by measuring the standard deviation of each observable across all integrations in the dataset and then dividing by the square root of the number of integrations. 
This gives the standard error of the mean, or the statistical uncertainty in the measurement of the mean observable. 
As discussed in Section \ref{sec:noise}, these errors may not dominate if calibration errors exist, which is often the case with AMI observables. 
For fits to calibrated data, following e.g. \citet{2013MNRAS.433.1718I}, we inflate error bars to match the scatter measured across the closing triangles (such that the reduced $\chi^2$ of the best-fit model is roughly equal to one). 

\subsection{Calibration}\label{sec:calib}
To calibrate the HIP 65426 observables, we subtract the (closure) phases of a reference from the HIP 65426 phases, and divide the (squared) visibilities of a reference into the HIP 65426 visibilities. 
Since these data consist of a single science pointing sandwiched between two reference star pointings, we test four different calculations of that reference. 
The first two are simply using each individual PSF star as the calibration reference. 
We calculate a third reference by fitting a linear trend in time to the two PSF star observables for each baseline / closing triangle, and then sampling that trend at the time of the HIP 65426 observable.
Lastly, we calculate a calibration reference by averaging (with equal weighting) the observables of the two reference stars. 
As discussed in Section \ref{sec:fobs}, we find that the calibration quality is maximized when only the HD 116084 PSF star is used as the calibration reference. 

\subsection{Modeling and Contrast Curve Generation}
\citet{ray_subm} describes the procedure we apply for model fitting, and the results of model fits to the calibrated HIP 65426 dataset.
For this paper we focus on contrast curves as one method of quantifying NIRISS AMI performance. 
To measure achievable contrast, following \citet[][]{2019JATIS...5a8001S}, we fit a grid of single-companion models to the data, varying their separation, position angle, and contrast. 
We average over position angle, resulting in a grid in just separation and contrast. 
For each separation, we then calculate the 5$\sigma$ detectable contrast by finding the companion model with a $\chi^2$ that is greater than the null (no companion) model $\chi^2$ by $\Delta\chi^2=25$.
A $\Delta\chi^2$ interval of 25 (corresponding to 5$\sigma$ for a distribution with one degree of freedom) is used to calculate the achievable contrast since, for each separation, only one parameter (contrast) determines the difference between each model and the null model.

This procedure assumes that the reduced $\chi^2$ of the best-fit companion model is equal to one, and that the observables' error bars are Gaussian.
The first assumption is valid since we use the inflated error bars described in Section \ref{sec:ext}, which are constructed such that the best-fit reduced $\chi^2$ is roughly equal to one.
Additionally, to calculate the contrast curves we explicitly rescale the $\chi^2$ values so that the best-fit has a reduced $\chi^2$ of exactly one, (effectively assuming that any poor fit quality is caused by imperfect error bar estimation).
For the second assumption, as discussed in detail in Section \ref{sec:perf}, the AMI observables' uncertainties are dominated by calibration errors, which may not necessarily follow a Gaussian distribution.

While the possibility of non-Gaussian errors means that these contrast curves may be optimistic estimates for practical AMI datasets, we use them here for a few reasons. 
The first is that the distributions of calibrated closure phases and squared visibilities can be shown to be roughly Gaussian (albeit with a higher standard deviation than the statistical errors, which we capture in the error bar estimation described in Section \ref{sec:ext}).
Furthermore, self-calibrated datasets that can be used to explore potential performance will have calibration errors equal to zero, making their error bars better modeled by Gaussian distributions.
Lastly, the computational expediency of this method compared to e.g.~companion injections and recoveries also means that a large number of contrast curves can be generated quickly to enable the wide variety of calibration tests performed here.
However, we note that for an AMI dataset dominated by calibration errors, injection and recovery tests are useful both for exploring the validity of the $\Delta\chi^2$ approach and for determining robust scientific detection limits.

\section{Results: NIRISS AMI Performance and Noise Characteristics}\label{sec:perf}

\begin{figure*}
    \centering
    \includegraphics[width=0.8\textwidth]{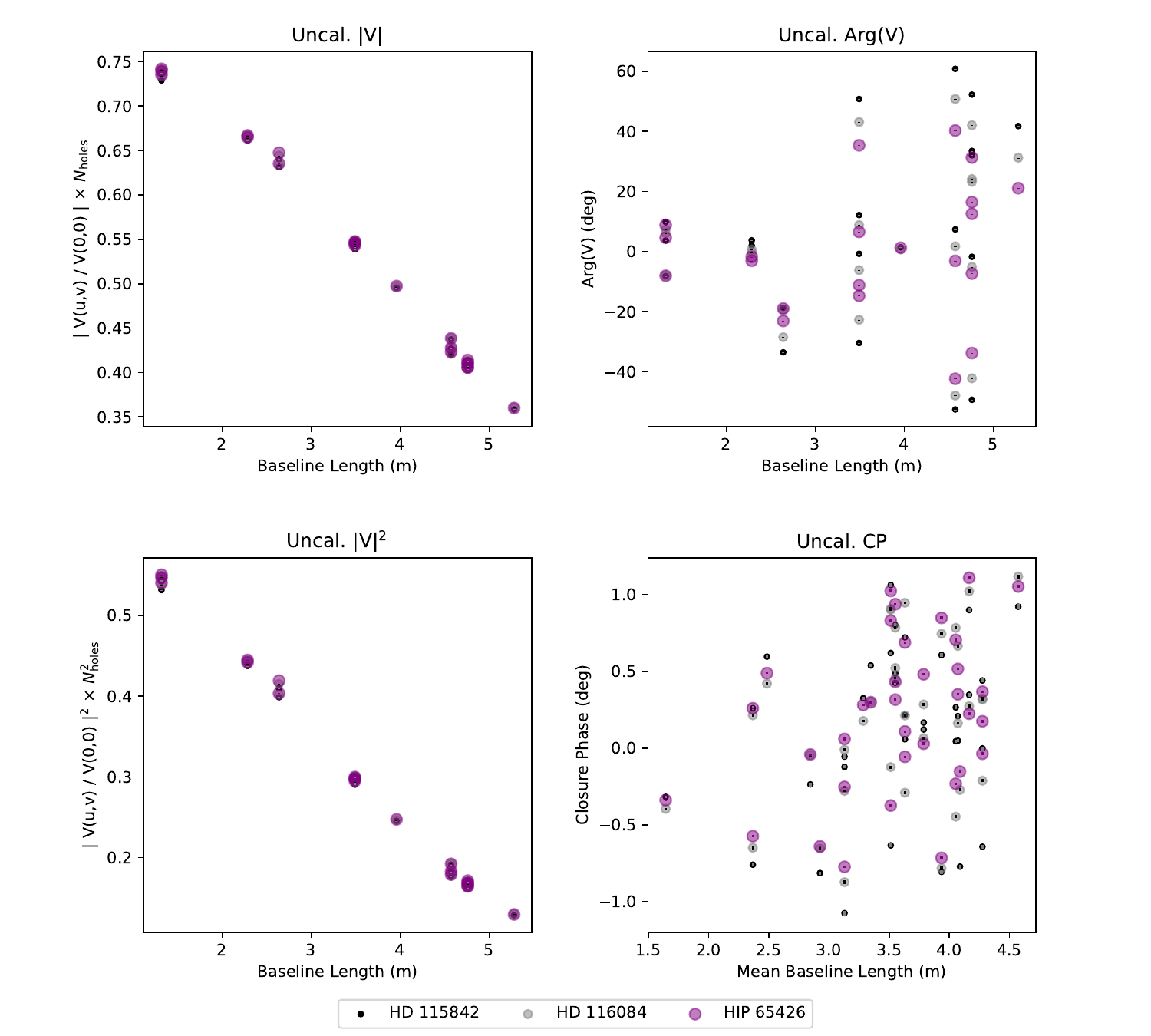}
    \caption{Uncalibrated Fourier observables. The small black points, medium grey points, and large purple points show observables for HD 115842, HD 116084, and HIP 65426, respectively. The top left panel shows complex visibility amplitudes versus baseline length, normalized by the total signal in the images and multiplied by the number of mask holes. These values should all be equal to 1 for a perfect-Strehl point source observation. The bottom left panel shows squared visibilities versus baseline length with the same normalization applied. The top right panel shows complex visibility phases versus baseline length, and the bottom right panel shows closure phases versus the average length of the three baselines in each triangle. In all panels statistical error bars (estimated from the standard error measured across all integrations) are smaller than the size of the symbols.}
    \label{fig:uncal}
\end{figure*}

\begin{figure*}
    \centering
    \includegraphics[width=0.9\textwidth]{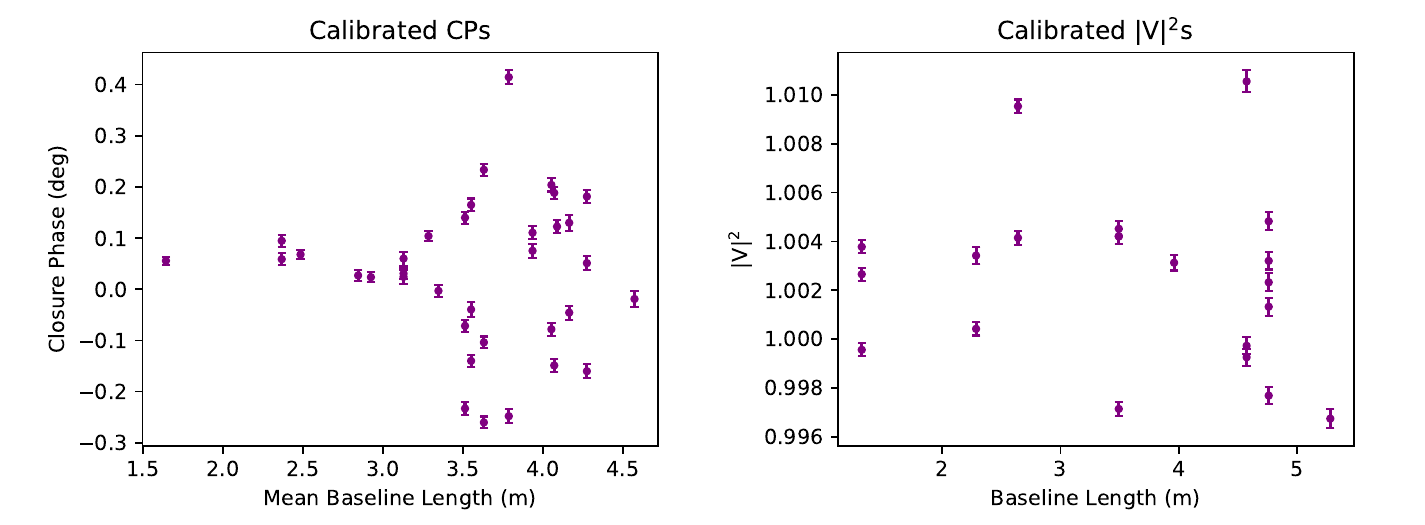}
    \caption{Calibrated HIP 65426 Fourier observables. The left panel shows the HIP 65426 closure phases versus the mean baseline length of each triangle, after calibrating with the HD 116084 reference. The right panel shows the HIP 65426 squared visibilities after calibrating with the HD 116084 reference. The error bars show the statistical errors that have been propagated taking the calibration procedure into account. }\label{fig:cal}
\end{figure*}

\subsection{Fully-Reduced Fourier Observables}\label{sec:fobs}
Figure \ref{fig:uncal} shows the uncalibrated Fourier observables for HIP 65426 and the two reference stars HD 115842 and HD 116084. 
In all four panels, the sizes of the statistical error bars are significantly smaller than the scatter in the observables. 
This shows that significant static errors exist, which require calibration for removal. 
Figure \ref{fig:cal} shows the best calibration of the squared visibilities and the closure phases, which is achieved when the HD 116084 observables are used as the calibration reference and when the 48-pixel-FWHM super-Gaussian window is applied to all images. 
We show just these two observables since they are used to generate the detection maps in \citet{ray_subm}. 
Comparing the calibrated data in Figure \ref{fig:cal} to the uncalibrated data in Figure \ref{fig:uncal} shows that the calibration significantly reduces the static errors in the data. 
In the following subsections we discuss the random, static, and calibration errors in more detail.

\begin{figure*}
    \centering
    \includegraphics[width=0.9\textwidth]{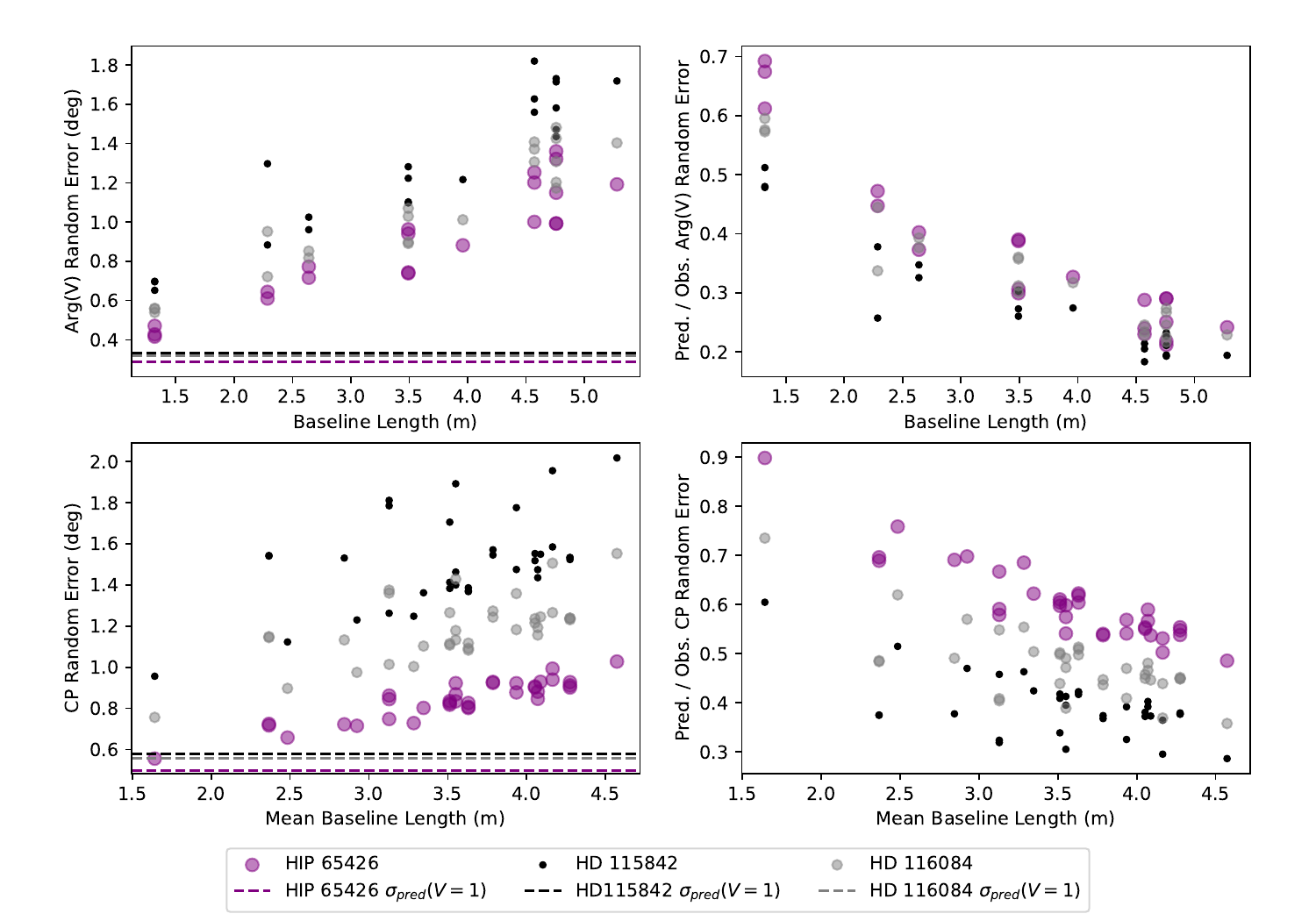}
    \caption{Left panels show the observed standard deviation of each visibility phase (top) and closure phase (bottom) versus baseline length. The dashed lines show the predicted scatter for phase observables with a fringe visibility equal to one. Right panels show the ratio between the predicted and observed scatter for the two phase quantities. The decrease in this ratio as a function of baseline length is consistent with a systematic decrease in fringe visibility. In all panels, the small black points, medium grey points, and large purple points show observables for HD 115842, HD 116084, and HIP 65426, respectively.}
    \label{fig:rand_vs_bl}
\end{figure*}

\subsection{Random Errors, Static Errors, and Fringe Visibility}\label{sec:randerrs}
The magnitude of the static phase errors can be roughly estimated by taking the standard deviation of the final, time-averaged phases.
For complex visibility phases this standard deviation is measured across the baseline index, and for the closure phases it is measured across the triangle index. 
These values are $\sim21^\circ-31^\circ$ for the complex visibility phases, with the scatter increasing from HD 115842 to HD 116084 to HIP 65426. 
The closure phase static errors range from $0.53^\circ-0.58^\circ$ for the three objects.
The fact that the errors are inflated for the phases compared to the closure phases is due to the self-calibrating properties of the closure phases. 

The statistical error bars between the phases and closure phases are similar, with values of $\sim0.008^\circ-0.012^\circ$ for the three objects. 
If these errors came only from the terms in Equations \ref{eq-pn} and \ref{eq-cp}, the closure phase errors should be approximately a factor of $\sqrt{3}$ higher than the phase errors. 
This suggests that there is an additional random noise term affecting the visibility phases and not the closure phases.
One possible explanation for this is pointing jitter, which would calibrate down in the closure phases (since it introduces a ramp in Fourier phase for each integration) but not in the phases. 
Jitter combined with flat-field variation would also contribute first-order phase errors to the visibility phases, but would only contribute at third-order and higher for closure phases.
A quantitative comparison of pointing jitter and flat field errors' effects on statistical phase and closure phase errors will be the subject of future work. 

The left panels of Figure \ref{fig:uncal} show that the static errors in the (squared) visibility amplitudes vary as a function of baseline length. 
The normalization shown in Figure \ref{fig:uncal} (dividing the amplitudes by the zero-spacing amplitude, and multiplying by the number of holes) is constructed so that a perfect-Strehl point source would have all visibilities equal to one. 
The dropoff as a function of baseline length suggests that the coherence decreases systematically for the longer baselines, which is a well-known phenomenon in ground-based AMI datasets \citep[e.g.][]{1999PhDT........19M,2021AJ....161...28S}.

The drop in raw visibility amplitudes is relevant for the random noise predictions given by Equations \ref{eq-pn} and \ref{eq-cp}, since the visibilities map directly to the $V$ terms in their denominators. 
This implies that the statistical errors of the (closure) phases should increase as a function of the (mean) baseline length. 
Figure \ref{fig:rand_vs_bl} shows that this is the case. 
The random variations of each complex visibility phase and each closure phase increase as a function of baseline length. 

If all count-rate quantities in Equation \ref{eq-pn} were properly estimated, and all random noise sources were taken into account, the ratio between the random complex visibility phase errors and the Equation \ref{eq-pn} predictions for 100\% Strehl should equal the fringe visibility for each baseline. 
These quantities are shown in the top right panel of Figure \ref{fig:rand_vs_bl}, and they show a similar characteristic decrease as the raw visibilities shown in Figure \ref{fig:uncal}. 
However, two important differences exist between the two.
The first is that the raw visibility amplitudes are always higher than the visibilities that would be inferred from the ratios. 
This indicates that there are additional random noise sources for all three objects that are not captured by Equation \ref{eq-pn}.
The second is that there are variations in the ratios from object to object that are not seen in the raw visibility amplitudes.
This indicates that the additional random noise sources are more severe for the two calibrator stars than for HIP 65426, and most severe for the HD 115842 calibrator. 
We explore whether this could stem from the number of groups used for these two objects in Section \ref{sec:selfcal}. 

The tests in Figure \ref{fig:rand_vs_bl} and the raw visibilities in Figure \ref{fig:uncal} suggest that we should not be adopting $V=1$ in Equations \ref{eq-pn} and \ref{eq-cp}.
However, adopting the raw visibility values in Figure \ref{fig:uncal} causes the photon noise predictions from Equation \ref{eq-cp} to be larger than the observed scatter about the mean closure phases for all three objects.
This may be due to the fact that both windowing and averaging over multiple pixels in the Fourier transform can smooth out noise variations in a way that is not captured by Equations \ref{eq-pn} and \ref{eq-cp}.
Since this is the case, and since AMI observers often assume $V=1$ during expected signal to noise calculations, in the closure phase noise comparisons that follow we assume $V=1$ for theoretical noise calculations.
This can be thought of as a conservative choice for comparing measured performance to theoretical predictions, since assuming $V=1$ results in the lowest predicted closure phase scatter and thus the worst underperformance relative to theoretical predictions.
We note that an intermediate assumption that takes into account the slope of the visibilities versus baseline (but not the overall normalization) may be a more accurate predictor of the expected closure phase noise levels.

\subsection{Reference PSF Calibration Quality}\label{sec:cal_qual}
Here we quantify the closure phase calibration quality by making comparisons to the theoretical predictions of Equation \ref{eq-cp} assuming $V=1$. 
With a large enough number of collected photons, static errors will dominate over random errors. 
The observables can thus be expected to have a $\sigma\propto\sqrt{n_{int}}$ trend, where $n_{int}$ is the number of integrations, until the systematic noise floor is reached. 
With perfect calibration, all underlying systematics would be removed, and the calibrated data would follow the $\sigma\propto\sqrt{n_{int}}$ trend.

To assess the calibration quality, we compare the scatter in the calibrated HIP 65426 data to expectations for photon noise. 
For this test and in some of the analysis that follows, we use jackknife resampling \citep[e.g.][]{10.1214/aoms/1177729989} to create shallower datasets by systematically removing individual integrations.
For example, to calculate the calibrated closure phase scatter in a dataset with $n_{int}$ integrations, we create a large number of ``jackknifed" datasets by selecting different subsets of $n_{int}$ integrations pulled from the entire dataset for a given object. 
For each subset of $n_{int}$ integrations, we calculate the mean set of 35 closure phases. 
We then perform calibration as described in Section \ref{sec:calib}, and calculate the standard deviation of the 35 calibrated closure phases.
We take the average of that standard deviation across all jackknifed $n_{int}$-integration datasets as representative of the calibrated closure phase scatter.

We apply this jackknife resampling procedure to the HIP 65426 integrations, and calibrate using all of the integrations from the HD 116084 PSF calibrator.
We use HD 116084 since it provided better calibration quality (final closure phase scatter of $0.14^\circ$).
We split the HIP 65426 data up into shallower datasets with depths ranging from $n_{int} = 2$ integrations to the total number of $n_{int} = 10950$.
For each dataset we then subtract the average of all of the HD 116084 integrations to calibrate. 
This creates a large number of HIP 65426 datasets with varying depths that have all been calibrated with the highest signal to noise HD 116084 measurements. 
We calculate the scatter for each set of calibrated closure phases by taking the standard deviation across the different triangles. 

\begin{figure*}
    \centering
    \includegraphics[width=\columnwidth]{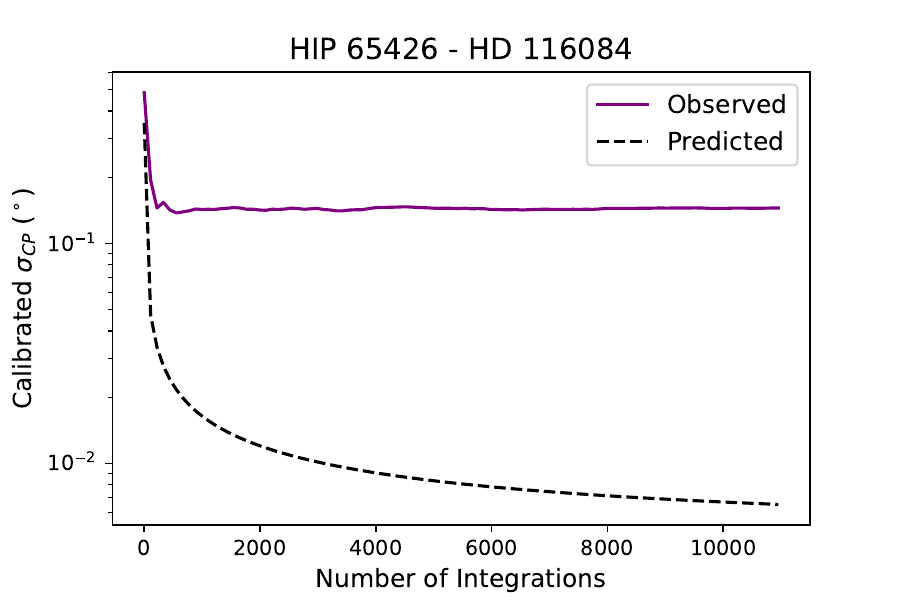}
    \includegraphics[width=\columnwidth]{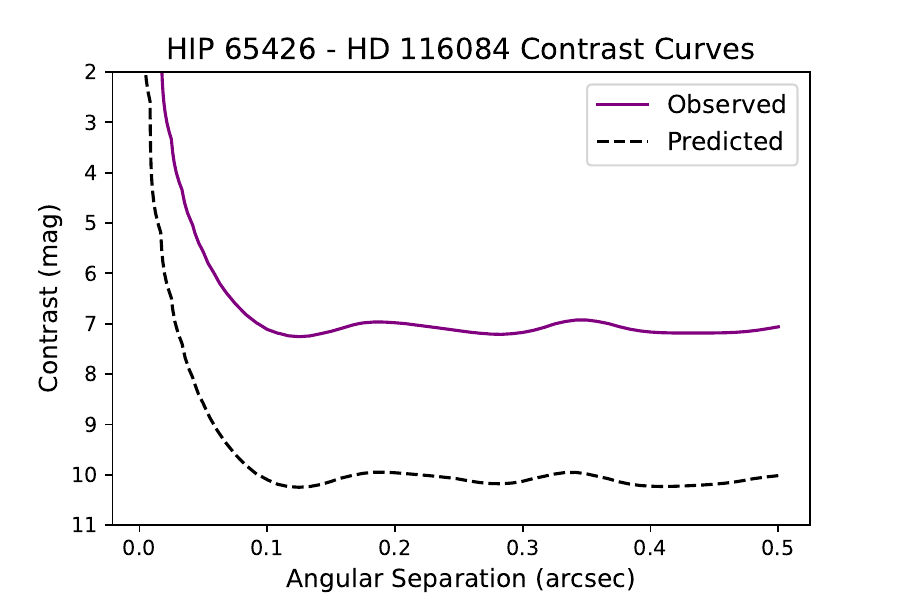}
    \caption{Left: Comparison between HIP 65426 calibration quality and expectations for photon noise. The purple solid line shows the scatter in the time-averaged calibrated HIP 65426 closure phases, as a function of the number of integrations that were included in the average. The dashed black line shows the expected closure phase scatter for photon noise. The noise in the HIP 65426 closure phases calculated using all 10950 integrations is inflated by a factor of 22.25 compared to photon noise expectations. Right: Contrast curves calculated from the observed HIP 65426 observations (using all 10950 integrations) calibrated with the entire HD 116084 dataset, compared to theoretical predictions. The purple solid line shows the observed contrast, and the black dashed line shows the predicted contrast. The observations underperform compared to predictions by a factor of $\sim3$ magnitudes.}
    \label{fig:sci-ref2-pncomp}
\end{figure*}

To calculate expectations for the noise floor, we calculate the total counts in the subframed, super-Gaussian windowed images. 
We convert to the total number of collected photons accounting for the exposure parameters ($t_{int} = 0.22632$ for HD 116084, $t_{int} = 0.98072$ for HIP 65426) and the NIRISS gain (1.61 e$^-$/ADU). 
This results in 9.87e5 photons per integration for HIP 65426 (for a total of 1.08e10 photons for all 10950 integrations), and 7.99e5 photons per integration for HD 116084 (for a total of 1.28e10 photons for all 16000 integrations).
We then use Equation \ref{eq-cp} to predict the closure phase scatter in the photon-noise limit.  
For each HIP 65426 dataset consisting of $n_{int}$ averaged integrations, we also calculate the ratio between the resulting scatter and the predicted scatter given the total number of photons for those $n_{int}$ integrations.

Figure \ref{fig:sci-ref2-pncomp} shows the results. 
The scatter in the calibrated HIP 65426 closure phases decreases to a depth of about 200 integrations, and then flattens out at $\sim0.14^\circ$. 
This shows that there are residual systematic errors present in the calibrated data, at a level consistent with expectations for $\sim$2e8 photons. 
The ratio between the observed scatter and the predicted scatter for the full dataset depth ($n_{int}=10950$) is 22.25. 
This amounts to a degradation in achievable contrast of $\sim$3 mag. 

As a consistency check, we performed the same tests as above using HD 115842 as a reference for HIP 65426, and using HD 116084 as a reference for HD 115842. 
When the full HD 115842 dataset is used as a PSF reference for HIP 65426, the calibrated closure phase scatter reaches a level of $\sim$0.26$^\circ$ by the same depth of $\sim$200 integrations. 
When HD 116084 is used as a PSF reference for HD 115842, the calibrated scatter reaches $\sim$0.25$^\circ$ by roughly the same number of integrations. 
These tests, especially the $\sim2\times$ higher scatter in the HIP 65426 data calibrated with HD 115842, show that varying static errors are present across the three datasets. 
This is also evidenced by the differing single companion fit results for the HIP 65426 dataset calibrated with each reference PSF \citep{ray_subm}.

\begin{figure*}
    \centering
    \includegraphics[width=\columnwidth]{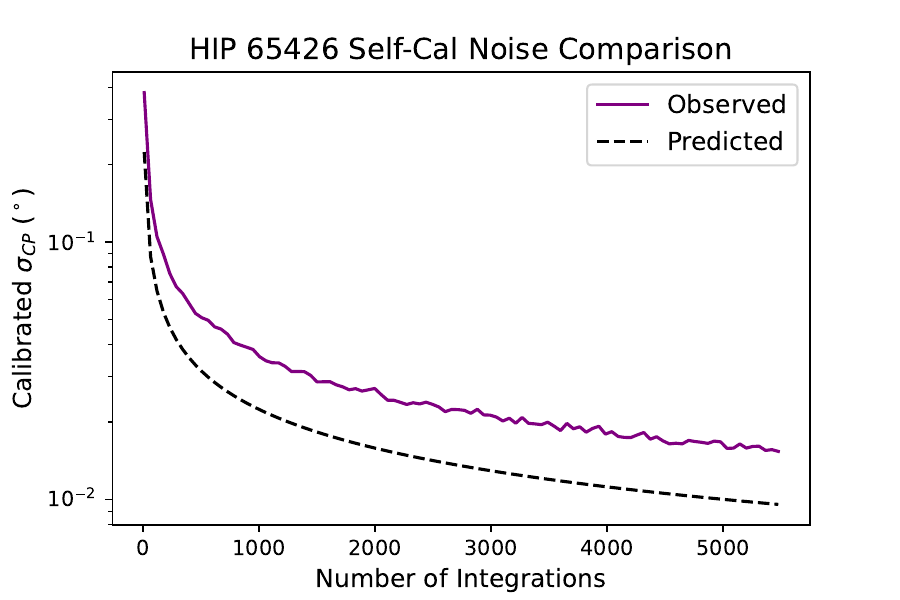}
    \includegraphics[width=\columnwidth]{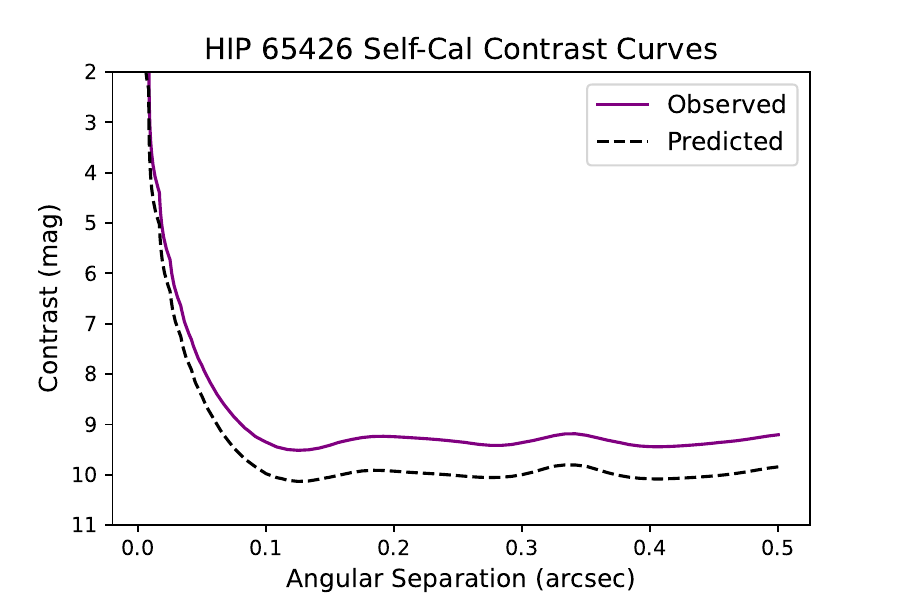}
    \includegraphics[width=\columnwidth]{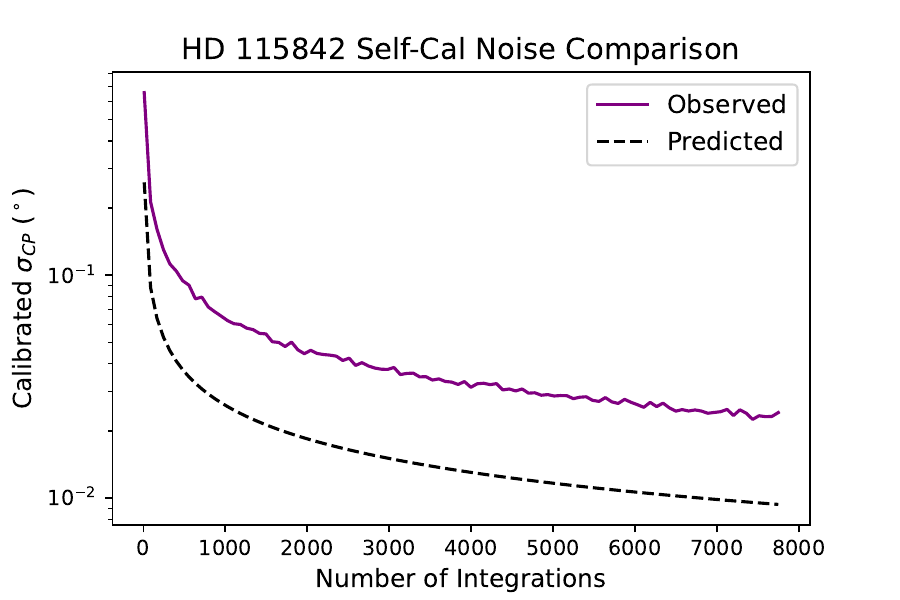}
    \includegraphics[width=\columnwidth]{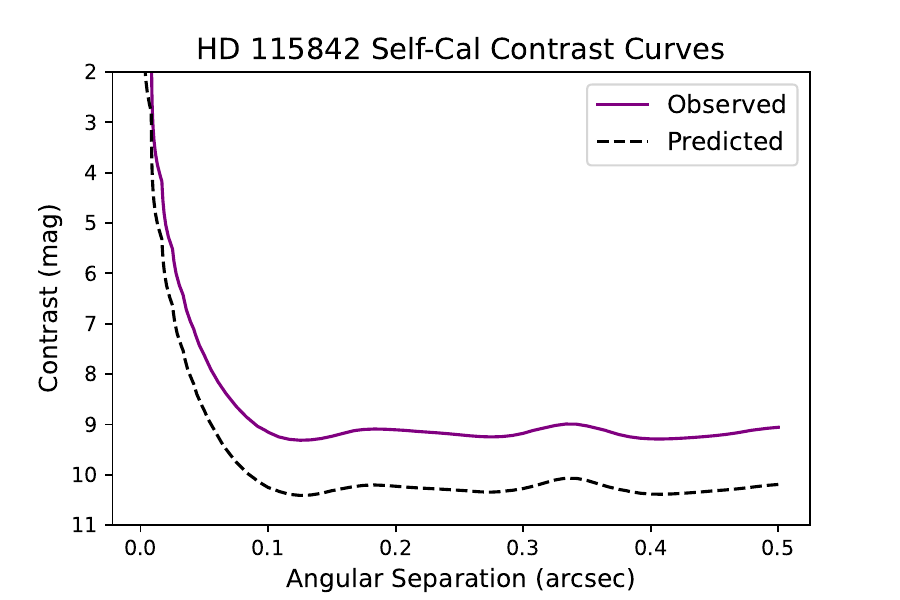}
    \includegraphics[width=\columnwidth]{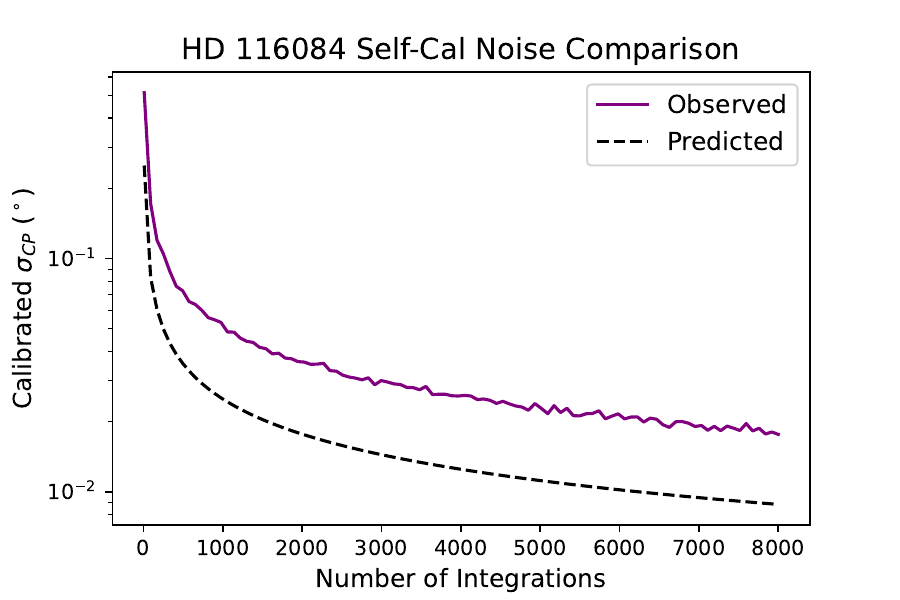}
    \includegraphics[width=\columnwidth]{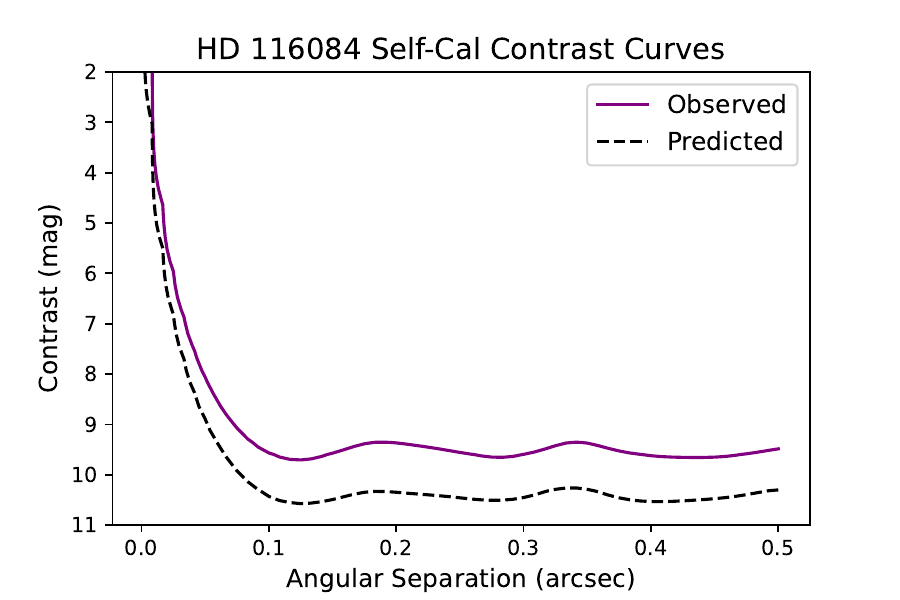}
    \caption{Left panels: Comparisons between self-calibration quality and expectations for photon noise, for HIP 65426, HD 115842, and HD 116084 from top to bottom. In each panel, the purple solid line shows the scatter in the time-averaged calibrated closure phases for a jackknifed dataset, as a function of the number of integrations that were included in the average. The dashed black line shows the expected closure phase scatter for photon noise. The ratio of these two quantities (observed / predicted) is steady for all three objects, at $\sim1.6$ for HIP 65426, $\sim2.5$ for HD 115842, and $\sim2.1$ for HD 116084. Right panels: Contrast curves calculated from the self-calibrated observations using the maximum number of integrations in each jackknifed dataset, compared to theoretical predictions. Top to bottom panels show HIP 65426, HD 115842, and HD 116084. In each panel, the purple solid line shows the observed contrast, and the black dashed line shows the predicted contrast.}
    \label{fig:selfcal-pncomp}
\end{figure*}

\subsection{Self-Calibration Tests: 
 Achievable Contrast with Perfect Calibration}\label{sec:selfcal}
We use ``self-calibration" tests to explore the source of these inflated errors and resulting lower contrast. 
To check that the underlying problem is indeed an uncalibrated static error, we jackknife each dataset into pairs of shallower datasets and use those pairs to self calibrate. 
We then perform the same tests as described in Section \ref{sec:cal_qual}.
For example, for HIP 65426, we create pairs of datasets with total numbers of integrations between 2 and 5475 (half the number of integrations of the complete HIP 65426 dataset). 
We average each of the two datasets to calculate two sets of mean closure phases, which we then calibrate against each other. 
We note that this test is only useful for characterizing the random noise level of the dataset, since the self-calibration process removes both the static errors and any science signal present in the data. 

Following self-calibration, we compare the results to photon noise following the procedure in Section \ref{sec:cal_qual}.
Figure \ref{fig:selfcal-pncomp} shows the results. 
For all three objects, the scatter in the self-calibrated dataset decreases proportionally to $\sqrt{n_{int}}$, where $n_{int}$ is the number of integrations included in each jackknifed dataset. 
The calibration quality compared to photon noise varies from object to object, with HIP 65426 reaching a noise level that is $\sim1.6\times$ the noise floor, HD 115842 at $\sim2.5\times$, and HD 116084 at $\sim2.1\times$.

\begin{figure}
    \centering
    \includegraphics[width=\columnwidth]{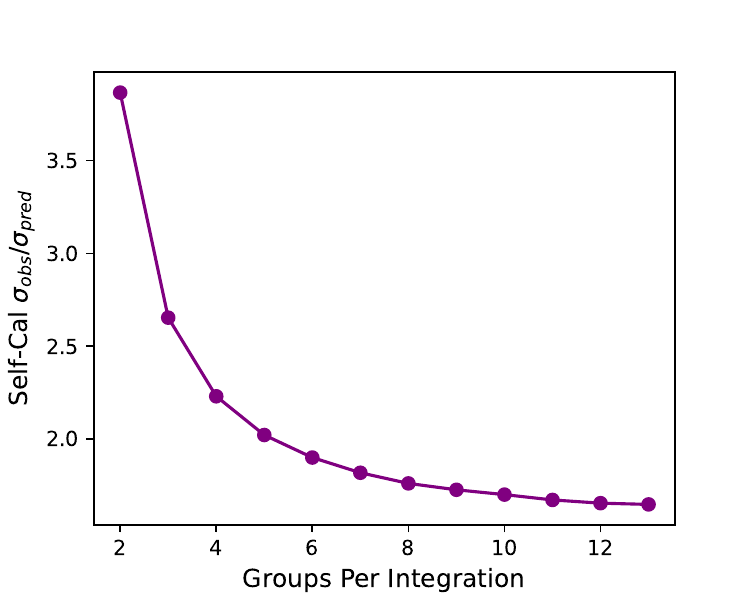}
    \caption{The ratio between the observed closure phase scatter and the theoretical noise floor for self-calibrated HIP 65426 datasets, as a function of the number of groups included per integration. In this test, the jackknifed datasets have the maximum number of integrations (5475, half of the HIP 65426 total integrations). The calculation of the statistical noise floor includes both the higher read noise and lower photon count when fewer groups are used.}
    \label{fig:selfcal_groups}
\end{figure}

We note that, given the visibility caveats discussed at the end of Section \ref{sec:randerrs}, the overall normalization of these values is likely overly conservative, but their relative values indicate increased random errors for the two reference PSFs. 
In order to explore the source of this additional noise in a more controlled way, we calculate the scatter in the self-calibrated HIP 65426 data after varying the number of groups used per integration. 
As shown in Figure \ref{fig:selfcal_groups}, the ratio between the observed scatter and the noise floor is worse when fewer groups are used. 
This suggests that the additional noise may be a detector-level effect that is somewhat mitigated by increasing the number of reads up the ramp.  

Figure \ref{fig:selfcal_groups} also shows that the ratio between the observed scatter and the noise floor for the self-calibrated HIP 65426 data is higher than that for the calibrators when identical numbers of groups are considered.
The ratio for HIP 65426 is $\sim 3.9$ for a 2-group reduction, and $\sim2.7$ for a 3-group reduction, compared to 2.5 for the 2-group HD 115842 dataset and 2.1 for the 3-group HD 116084 dataset.
This is also consistent with the additional noise being a changing detector-level effect such as time-variable pattern noise (visible as the vertical stripes in Figure \ref{fig:badpix}, which change from image to image).
Images with fewer collected photons will suffer more from these time-variable detector systematics, which are independent of the target brightness.
This is likely the cause of the inflated noise for the 2- and 3-group reductions of HIP 65426, which is significantly fainter than the two calibrators.

These self calibration tests show that, with adequate removal of static errors, NIRISS can reach unprecedented broadband AMI contrasts despite the slightly inflated random errors for the few-group integrations. 
This is illustrated in the right panels of Figure \ref{fig:selfcal-pncomp}. 
These show contrast curves for the deepest datasets in the left panels, where the two jackknifed datasets consist of half the total number of integrations for each object.
The achievable contrasts at separations $\gtrsim\lambda/D$ range from 9-10 magnitudes. 
The self-calibrated HIP 65426 contrast, at $\sim9.5$ mag, outperforms the final HIP 65426 reduction (calibrated with HD 116084) by $\sim 2.5$ mag. 
This motivates the need to better understand and calibrate the residual errors shown in Figure \ref{fig:sci-ref2-pncomp}.

\begin{figure*}
    \centering
    \includegraphics[width=\textwidth]{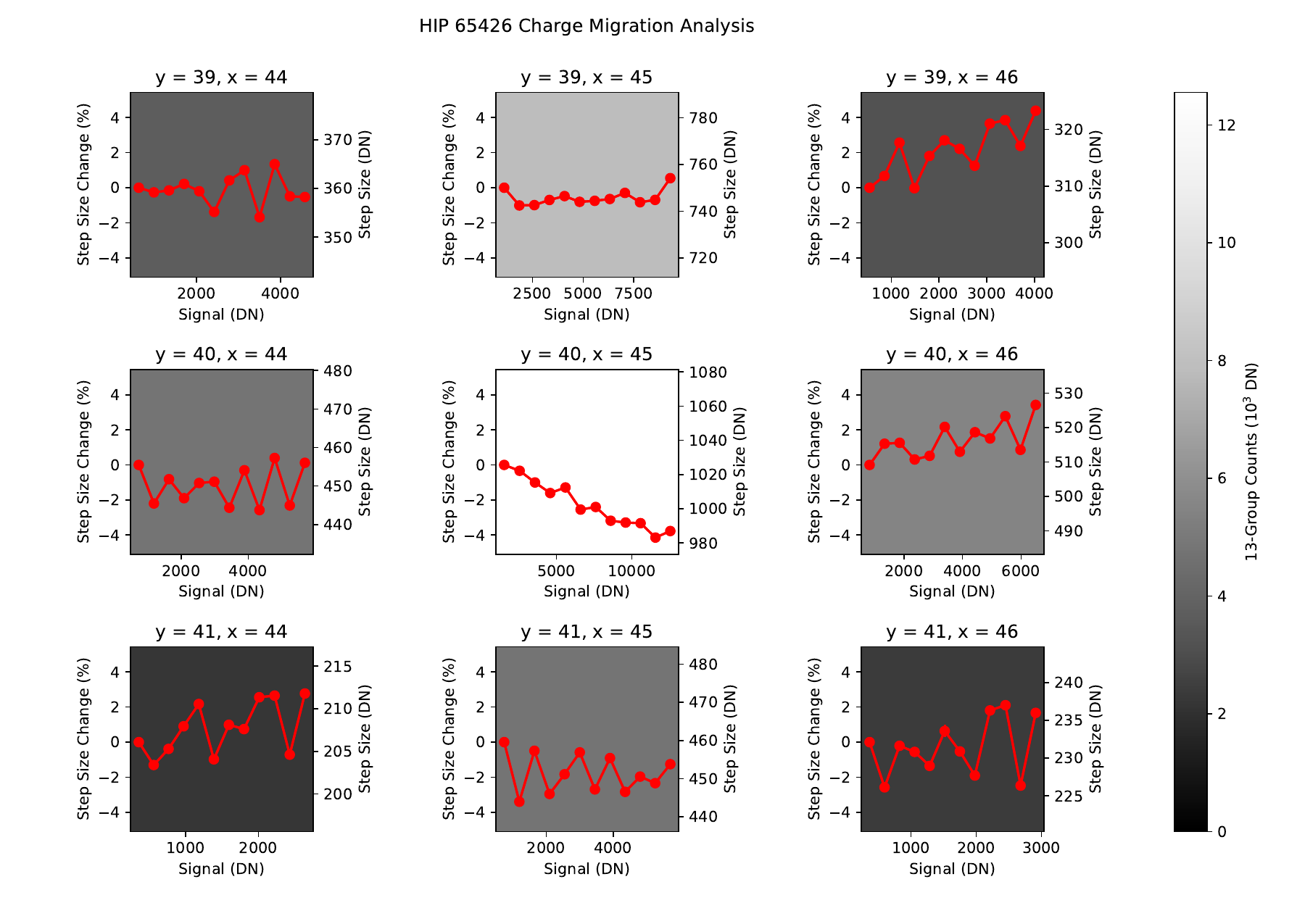}
    \caption{Charge migration analysis for HIP 65426. Each panel corresponds to a single pixel, with the nine panels centered around the brightest pixel in the AMI PSF. In each panel, the red line shows the change in accumulated charge relative to the first group pair (left axis) as well as the absolute amount of charge accumulated by each group pair (right axis). The greyscale indicates the maximum accumulated charge in each pixel (i.e. measured by the final two group pairs), with the colorscale set by the peak counts in the central, brightest pixel in the PSF. Comparing the lines in the various panels shows that the charge accumulation in the central pixel decreases by $\sim4\%$ over the course of an integration, while the charge accumulation in the pixels right, and diagonally right of center increases by $\sim4\%$. This charge migration causes signal-dependent PSF changes.}
    \label{fig:hip65426_cm}
\end{figure*}

\begin{figure*}
    \centering
    \includegraphics[width=\textwidth]{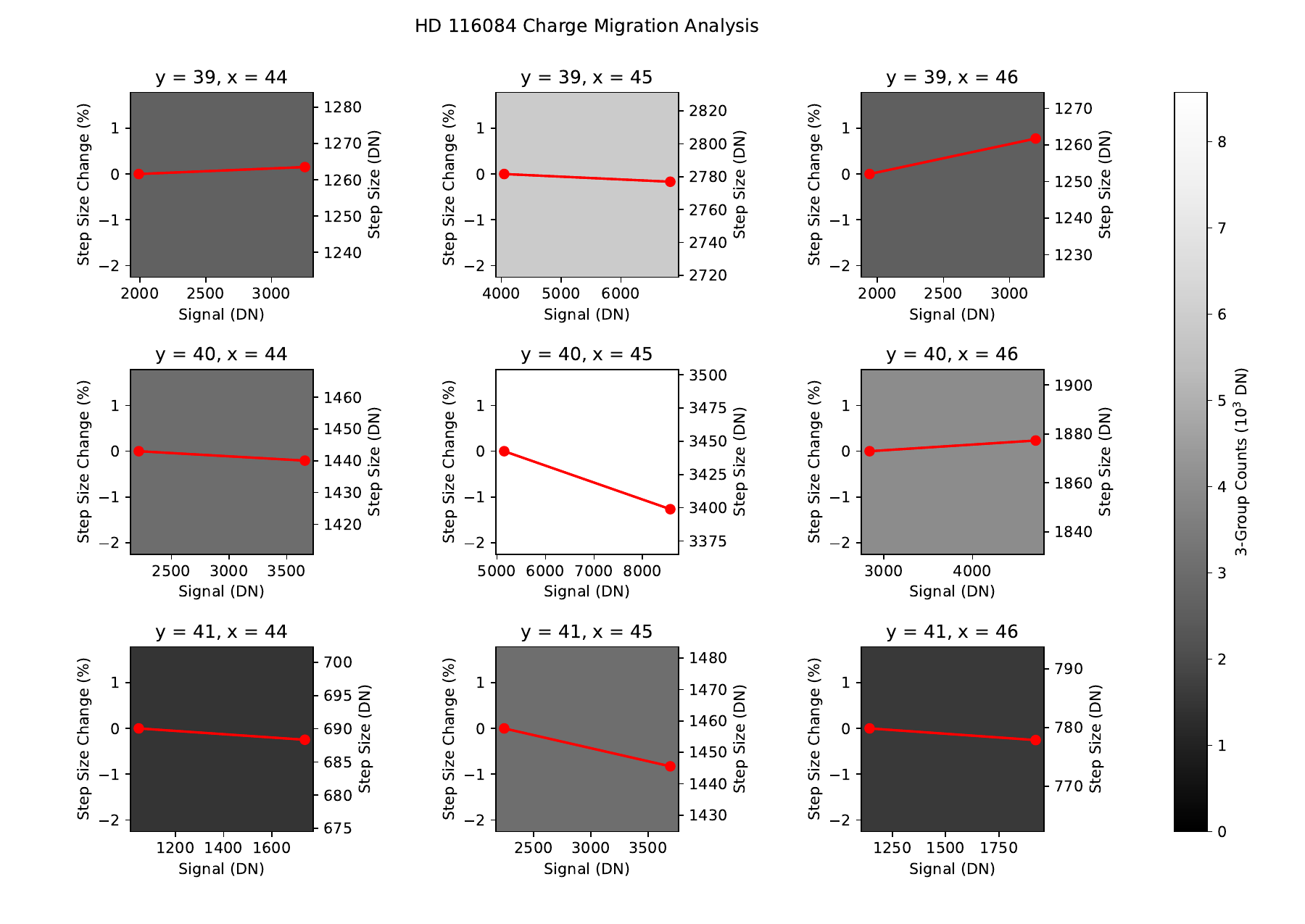}
    \caption{Charge migration analysis for HD 116084. Each panel corresponds to a single pixel, with the nine panels centered around the brightest pixel in the AMI PSF. In each panel, the red line shows the change in accumulated charge relative to the first group pair (left axis) as well as the absolute amount of charge accumulated by each group pair (right axis). The greyscale indicates the maximum accumulated charge in each pixel (i.e. measured by the final two group pairs), with the colorscale set by the peak counts in the central, brightest pixel in the PSF. Comparing the lines in the various panels shows that the charge accumulation in the central pixel decreases by $\sim1\%$ over the course of an integration, while the charge accumulation in the pixels right, and diagonally right of center increases by $\sim1\%$. This charge migration causes signal-dependent PSF changes.}
    \label{fig:hd116084_cm}
\end{figure*}

\subsection{Charge Migration Levels and PSF Effects}
One possible source of systematic differences in PSF shape is charge migration (CM).
This is a physical phenomenon where the electric field generated by accumulated charges in one location on the detector leads to the deflection of newly incident charges onto a slightly different location on the detector \citep[e.g.][]{2018PASP..130f5004P,2020PASP..132a4501H}.
These charge deflections cause the PSF to smear out spatially, with charges from the brightest pixels ``spilling over" onto adjacent ones.
The physical mechanism behind charge migration means that it becomes more dramatic at higher levels of total accumulated charge, causing the PSFs for brighter objects to smear out more. 
These brightness-dependent PSF changes are often referred to as the ``brighter-fatter effect."

H2RG detectors have been demonstrated in the laboratory to exhibit charge migration for high levels of accumulated charge \citep[$\sim20,000~\mathrm{e^{-}}$;][]{2018PASP..130f5004P}. 
This $\sim20,000~\mathrm{e}^{-}$ signal limit is generally used during \textit{JWST}/NIRISS observation planning, and the ERS 1386 observations were designed to be below this count level for individual integrations. 
While severe charge migration is thus unlikely, achieving exquisite contrast of $\sim9-10$ mag could be negatively affected by very low levels of CM. 
We thus analyze the \texttt{jwst} 4D data products to assess the degree of charge migration in the data for each object and to quantify its effect on their relative PSFs, following a procedure identical to that in \citet{2018PASP..130f5004P}. 

The 4D \texttt{jwst} data products have shapes $[n_{int}, n_{g}, n_y, n_x]$, where $n_{int}$ is the number of integrations, $n_g$ the number of groups, and $n_y$ and $n_x$ the number of pixels in the y and x directions. 
We start with the 4D datacubes that have had all standard reduction steps applied, including superbias, reference pixel, linearity, persistence, and dark current corrections. 
These would normally next undergo ramp fitting to produce rate images. 
For each integration, we take the differences between adjacent groups to calculate the total charge accumulated (in DN) at each step up the ramp. 
For any given pixel, we can then plot the accumulated charge at each step, as a function of the total accumulated signal at that step (calculated as the average of the two adjacent groups). 

Figure \ref{fig:hip65426_cm} shows the results for the central nine pixels of the 13-group observations of HIP 65426. 
This analysis shows that the charge accumulation in the central, brightest pixel (x=45, y=40) decreases slightly as the total signal increases from $\sim1500$ DN to $\sim13,000$ DN. 
The absolute step size decreases by $\sim40$ DN from $\sim1025$ DN to $\sim986$ DN.
These changes correspond to a fractional decrease relative to the first group pair of $\sim4\%$.
These decreases in the central pixel are accompanied by $\sim$15-20 DN (few percent level) increases in the pixels directly right (x=46, y=40), and up and right (x=46, y=39) of center.

Figure \ref{fig:hd116084_cm} shows the results for the same nine pixels in the 3-group observations of the HD 116084 PSF reference. 
In this case, a similar absolute decrease occurs for the central pixel, with step sizes decreasing from $\sim3442$ DN to $\sim3398$ DN as the total charge increases from $\sim5150$ DN to $\sim8570$ DN.
Since the individual steps up the ramp are larger, this decrease corresponds to a fractionally lower change in step size of $\sim1\%$. 
For these data, the greatest changes in adjacent pixels are a $\sim15$ DN ($\sim1\%$) decrease in the pixel directly below center (x=45, y=41), and a $\sim10$ DN ($\sim1\%$) increase in the pixel up and right of center (x=46,y=39). 
Since the HD 115842 observations only utilized two groups, we cannot investigate CM levels for that object.

These tests show that small amounts of charge move between pixels as integrations move up the ramp. 
To investigate the degree to which this CM affects the PSF shape, we compare the central nine pixels of the PSF for different reductions of the HIP 65426 dataset that include varying numbers of groups. 
Since AMI observations are sensitive to fractional changes in the PSF (i.e. overall normalization of the image does not change the closure phase signal), we compare the fluxes in the central nine pixels after normalizing by the central, brightest pixel. 
In this case, we use \texttt{jwst} to produce 3D \texttt{rateints} datacubes from subsets of the HIP~65426 4D cubes, varying the maximum number of groups ($n_{g,max}$) used for ramp fitting. 
We then average the central nine pixels over all integrations and normalize them by the signal in the central brightest pixel. 
We take differences between reductions with different $n_{g,max}$ values to quantify the fractional changes in the PSF cores. 

\begin{figure}
    \centering
    \includegraphics[width=0.8\columnwidth]{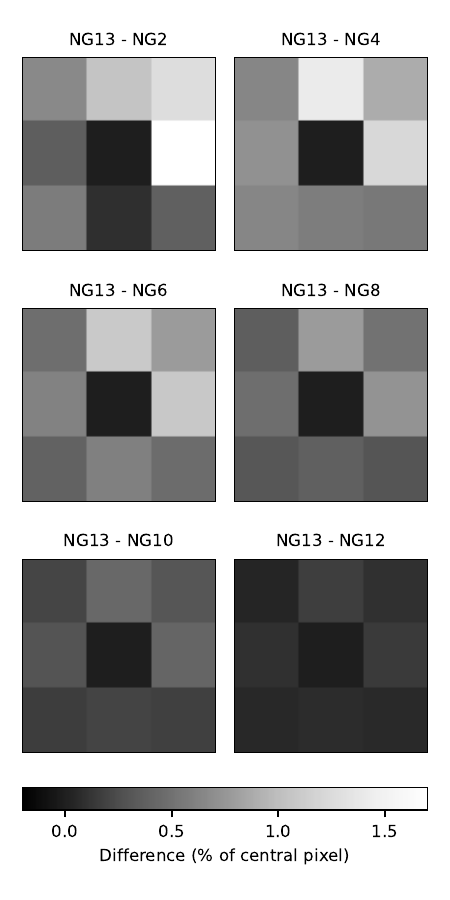}
    \caption{Changes in HIP 65426 PSF shape as a function of the maximum number of reads up the ramp. Individual panels show the difference between the central nine pixels in the average images for the full 13-group HIP 65426 reduction, and reductions with 2, 4, 6, 8, 10, and 12 maximum groups. All images are normalized to the central brightest pixel before differencing, in order to capture changes in the shape of the PSF. Comparison of the different panels shows that charge migration causes $\sim1\%$ level changes in the central brightest region of the PSF.}
    \label{fig:cm_psfs}
\end{figure}

Figure \ref{fig:cm_psfs} shows the differences between the $n_{g,max} = 2,~4,~6,~8,~10,~12$ reductions and the full $n_{g,max} = 13$ HIP 65426 reduction. 
The results show that the pixels up and to the right of center are $\sim0.7-1.7\%$ brighter in the $n_{g,max} = 13$ reduction compared to the reductions with fewer groups. 
This is consistent with the charge migration results in Figure \ref{fig:hip65426_cm}, which shows charge accumulation increasing in those pixels and decreasing in the central pixel as the reads up the ramp increase. 
The differences become smaller as the $n_{g,max}$ values approach 13, which is also consistent with this scenario since adjacent group pairs will have the most similar charge accumulation properties. 

\subsection{Charge Migration Systematic Errors and Self-Calibration Tests}\label{sec:cmcaltests}
We perform further self-calibration tests to explore whether the signal-dependent PSF changes in Figure \ref{fig:cm_psfs} could cause the calibration errors and contrast degradation in Section \ref{sec:cal_qual}.
We perform 11 different calibrations of the 13-group HIP 65426 dataset, each one using a HIP 65426 reduction with a smaller number of groups (2-12) to create a calibration reference. 
We then compare the resulting closure phase scatter to theoretical noise predictions, taking into account the smaller number of collected photons and higher read noise for the fewer-group HIP 65426 reductions when evaluating Equation \ref{eq-cp}.
We also calculate the achievable contrast of each calibrated HIP 65426 dataset, and the predicted achievable contrast. 
We note that comparing the self-calibration quality as a function of number of groups is valuable for isolating the effects of charge migration, since factors that normally change between a science target and PSF calibrator observation (such as target acquisition) are held constant.

\begin{figure*}
    \centering
    \includegraphics[width=\columnwidth]{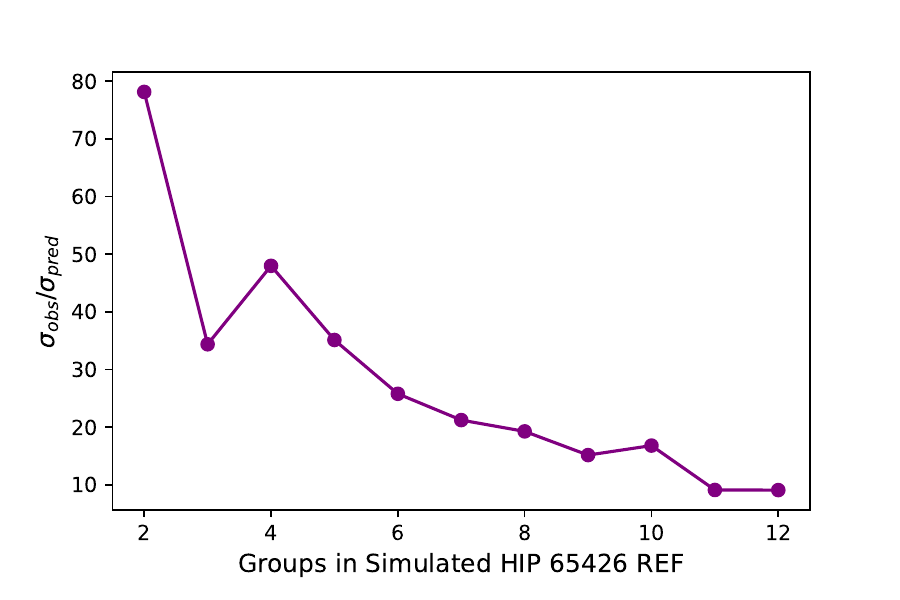}
        \includegraphics[width=\columnwidth]{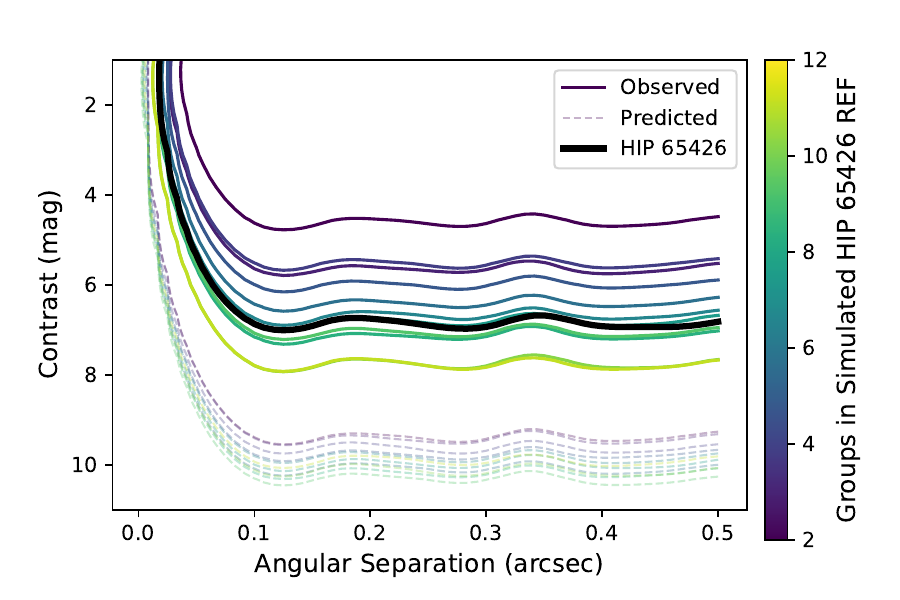}
    \caption{Self-calibration charge migration tests. The left panel shows the ratio between the observed closure phase scatter and the prediction for photon noise of the HIP 65426 13-group dataset calibrated using HIP 65426 datasets with varying numbers of groups. The right panel shows the resulting achievable contrast for each calibration (solid lines with colorscale), the predicted achievable contrast (dashed lines with colorscale), and the HIP 65426 contrast when calibrated with HD 116084 (thick black line).}
    \label{fig:cm_selfcal}
\end{figure*}

Figure \ref{fig:cm_selfcal} shows the results. 
As the number of groups in the calibration reference decreases, the ratio between the observed scatter and the theoretical prediction increases. 
The difference between the achievable contrast and the predicted contrast also becomes more drastic as the number of calibration groups decreases. 
The 9-group HIP 65426 reduction has the most similar peak counts to the HD 116084 reference star, at $\sim8597$ DN compared to $\sim8570$ DN. 
Calibrating the 13-group HIP 65426 dataset with its own 9-group reduction is thus a representative case for understanding the charge migration contributions when calibrating with HD 116084, since the severity of charge migration depends on the number of accumulated counts.
The contrast curve when the 9-group dataset is used as a reference is nearly identical to that observed for HIP 65426 calibrated with HD 116084, with a maximum achievable contrast of $\sim7.2$ mag compared to $\sim 7$ mag.

The 12-group calibration is also significantly worse than the 13-group self-calibration in Figure \ref{fig:selfcal-pncomp}, with an achievable contrast of $\sim7.6$ mag compared to $\sim9.5$ mag.
This makes sense given that a $\sim7.6$ mag companion at $\lambda/D$ would correspond to a $\sim0.1\%$ flux ratio at a distance of $<2$ pixels. 
The differences in pixel fractional fluxes between the 12- and 13-group reduction are comparable to this flux ratio at these separations (Figure \ref{fig:cm_psfs}).
The achievable contrast results when the fewer-group reductions are used as calibrators behave similarly to this case.
For example, the achievable contrast when the 2-group HIP 65426 reduction is used as a reference for the 13-group reduction, of $\sim4.5$ mag, is comparable to the fractional flux differences between those two reductions ($\gtrsim 1\%$) shown in Figure \ref{fig:cm_psfs}.

One caveat for interpreting Figure \ref{fig:cm_selfcal} is that, as shown in Figure \ref{fig:selfcal_groups}, the few-group reductions have inflated random errors compared to the predictions of Equation \ref{eq-cp}.
This implies that any degradation in calibration quality when few-group reductions of HIP 65426 are used as references could be partially caused by those inflated random errors.
However, even when this inflation is taken into account, the results in Figures \ref{fig:selfcal_groups} and \ref{fig:cm_selfcal} show that charge migration still dominates.
If there were no charge migration, then the ratio of the observed to predicted scatter for an $n_g$-group reference in Figure \ref{fig:cm_selfcal} should be equal to the 13-group ratio from Figure \ref{fig:selfcal_groups} added in quadrature to the $n_g$-group ratio from Figure \ref{fig:selfcal_groups}.
This is because Figure \ref{fig:selfcal_groups} isolates the increased random errors as a function of groups up the ramp, while Figure \ref{fig:cm_selfcal} includes contributions from both inflated random noise levels and calibration errors due to charge migration.
The fact that the noise ratios shown in Figure \ref{fig:cm_selfcal} are significantly higher than those in Figure \ref{fig:selfcal_groups} shows that charge migration is the dominant source of scatter in the calibrated closure phases for this test.

\section{Discussion}\label{sec:disc}

The statistical tests carried out here demonstrate that the under-performance of the HIP 65426 NIRISS AMI dataset (relative to photon noise with $V=1$) results from a combination of inflated random and calibration errors. 
The random visibility phase errors are larger than theoretical predictions, and increase systematically with baseline length, consistent with a drop in raw visibility amplitude. 
This suggests that the perfect-Strehl ($V$=1) assumption in predicting photon-noise-limited phase scatter is overly optimistic. 
However, this is not the only source of inflated random errors, since the relative inflation varies from object to object in this study (Figure \ref{fig:rand_vs_bl}). 
This may be due to increased levels of changing detector systematics for the objects where fewer numbers of groups were used (e.g. Figure \ref{fig:badpix}). 
Indeed, the behavior of the complex visibility phase scatter best matches the predictions given the raw visibility amplitudes for HIP 65426, where 13 groups were used per integration.

The self-calibration tests demonstrate that systematic differences between HIP 65426 and the PSF stars contribute significantly to the calibration errors in Figure \ref{fig:sci-ref2-pncomp}. 
While one obvious and exciting source of systematic PSF differences is intrinsic phase signals (such as companions), several lines of evidence show that this is not the case here. 
As described in \citet{ray_subm}, no global $\chi^2$ minimum exists for a single-companion fit to HIP 65426 calibrated with HD 116084.
This is also the case when HIP 65426 is calibrated with HD 115842, and the best-fit single-companion model has different parameters from the other calibration. 

Furthermore, neither HD 115842 nor HD 116084 had significant single companion signals in SPHERE AMI vetting data \citep{ray_subm}.
The NIRISS observations of the two objects calibrated against one another result in high closure phase scatter (Section \ref{sec:cal_qual}) with no global single-companion $\chi^2$ minimum. 
The fact that the reference stars do not calibrate well against each other also argues against spectral type mismatches as the source of the HIP 65426 calibration errors. 
HD 115842 and HD 116084 are spectral type B0.5 and B2, respectively, while HIP 65426 is spectral type A2. 
If spectral type mismatches were to blame, the scatter in the HIP 65426 closure phases calibrated with HD 116084 ($0.14^\circ$) should be higher than the scatter of the two reference stars calibrated against each other ($0.25^\circ$). 

The jackknife tests in Section \ref{sec:cmcaltests} show that differing amounts of charge migration can account for the HIP 65426 calibration errors (e.g. Figure \ref{fig:cm_selfcal}).
Differences of $\sim0.5-1\%$ in the central nine pixels of the PSF (compared to the peak pixel) lead to a factor of $\sim10-20$ inflation compared to photon noise expectations.
This produces achievable contrast nearly identical to that for HIP 65426 calibrated using HD 116084. 
Furthermore, the fact that the 9-group HIP 65426 dataset is just as bad a calibration reference as the 3-group HD 116084 dataset suggests that charge migration dominates over differing detector systematics associated with groups up the ramp. 
Indeed, the severity of the calibration errors when the 12-group HIP 65426 dataset is used as a reference may indicate that charge migration is the dominant noise source when PSF references are approximately within a spectral type and are observed close in time to the science target.

\section{Conclusions and Recommendations for Future NIRISS AMI Programs}\label{sec:conc}

We conducted a systematic investigation of the noise properties and performance of the \textit{JWST}/NIRISS aperture masking interferometry mode in the F380M filter. 
The results showed that both the random and calibration errors are inflated compared to photon noise predictions. 
The random error inflation as a function of baseline length also showed that theoretical noise predictions assuming $V=1$ are overly optimistic.
Furthermore, the self-calibration tests demonstrated that differences in charge migration between the science target and PSF references can account for the observed calibration errors and achievable contrast in the HIP 65426 dataset.

Based on this study, and in the absence of sophisticated charge migration calibration strategies, we make the following observation planning recommendations for future NIRISS AMI programs:
\begin{enumerate}[leftmargin=*]
\itemsep0em 
    \item To mitigate inflated random phase errors that may be caused in part by detector systematics we recommend maximizing the groups per integration. 
    \item To mitigate the effects of charge migration, we recommend observing PSF calibrators to a similar well depth as the science target.
    \item When possible, we recommend selecting PSF calibrators with similar brightnesses to the science target. This would allow for both the peak counts and number of groups to be matched as well as possible between the two datasets. 
\end{enumerate}

The above recommendations may produce overly stringent requirements on PSF calibrator stars for some science targets.
This motivates future studies to quantitatively explore the relative importance of factors such as differing detector readout settings, charge migration levels, calibrator spectral types, and thermal drifts between science targets and calibrators.
It also demonstrates the importance of developing sophisticated calibration strategies that can account for such differences between science targets and calibrators. 
These calibration strategies could be applied in pre-processing steps, such as including a PSF correction taking into account charge migration effects.
With enough archival calibrator datasets, they could also be applied directly to the Fourier observables after building a calibration reference from a library of observations \citep[e.g.][]{2019MNRAS.486..639K}. 
Such future calibration programs and analyses will work toward the exquisite contrast levels expected from NIRISS AMI, enabling unprecedented thermal-infrared exoplanet science at small angular separations.

\section*{Acknowledgements}
This work is based on observations made with the NASA/ESA/CSA James Webb Space Telescope. We are truly grateful for the countless hours that thousands of people have devoted to the design, construction, and commissioning of \textit{JWST}. This project was supported by a grant from STScI (JWST-ERS-01386) under NASA contract NAS5-03127. M.D.F.~is supported by an NSF Astronomy and Astrophysics Postdoctoral Fellowship under award AST-2303911. A.Z.G.~acknowledges support from Contract No. 80GSFC21R0032 with the National Aeronautics and Space Administration. M.B.~received funding from the European Union’s Horizon 2020 research and innovation programme under grant agreement No. 951815 (AtLAST). S.M.~is supported by a Royal Society University Research Fellowship (URF-R1-221669). R.A.M.~is supported by the National Science Foundation MPS-Ascend Postdoctoral Research Fellowship under Grant No. 2213312. J.M.V.~acknowledges support from a Royal Society - Science Foundation Ireland University Research Fellowship (URF$\backslash$1$\backslash$221932). A.Z. acknowledges support from ANID -- Millennium Science Initiative Program -- Center Code NCN2021\_080

\bibliography{references2}{}

\begin{thebibliography}{}
\expandafter\ifx\csname natexlab\endcsname\relax\def\natexlab#1{#1}\fi
\providecommand{\url}[1]{\href{#1}{#1}}
\providecommand{\dodoi}[1]{doi:~\href{http://doi.org/#1}{\nolinkurl{#1}}}
\providecommand{\doeprint}[1]{\href{http://ascl.net/#1}{\nolinkurl{http://ascl.net/#1}}}
\providecommand{\doarXiv}[1]{\href{https://arxiv.org/abs/#1}{\nolinkurl{https://arxiv.org/abs/#1}}}

\bibitem[{{Baldwin} {et~al.}(1986){Baldwin}, {Haniff}, {Mackay}, \&
  {Warner}}]{1986Natur.320..595B}
{Baldwin}, J.~E., {Haniff}, C.~A., {Mackay}, C.~D., \& {Warner}, P.~J. 1986,
  \nat, 320, 595, \dodoi{10.1038/320595a0}

\bibitem[{{Biller} {et~al.}(2012){Biller}, {Lacour}, {Juh{\'a}sz}, {Benisty},
  {Chauvin}, {Olofsson}, {Pott}, {M{\"u}ller}, {Sicilia-Aguilar}, {Bonnefoy},
  {Tuthill}, {Thebault}, {Henning}, \& {Crida}}]{2012ApJ...753L..38B}
{Biller}, B., {Lacour}, S., {Juh{\'a}sz}, A., {et~al.} 2012, \apjl, 753, L38,
  \dodoi{10.1088/2041-8205/753/2/L38}

\bibitem[{Bushouse {et~al.}(2022)Bushouse, Eisenhamer, Dencheva, Davies,
  Greenfield, Morrison, Hodge, Simon, Grumm, Droettboom, Slavich, Sosey, Pauly,
  Miller, Jedrzejewski, Hack, Davis, Crawford, Law, Gordon, Regan, Cara,
  MacDonald, Bradley, Shanahan, Jamieson, Teodoro, \&
  Williams}]{bushouse_2022_7059013}
Bushouse, H., Eisenhamer, J., Dencheva, N., {et~al.} 2022, JWST Calibration
  Pipeline, 1.7.0,  Zenodo, \dodoi{10.5281/zenodo.7059013}

\bibitem[{{Gardner} {et~al.}(2006){Gardner}, {Mather}, {Clampin}, {Doyon},
  {Greenhouse}, {Hammel}, {Hutchings}, {Jakobsen}, {Lilly}, {Long}, {Lunine},
  {McCaughrean}, {Mountain}, {Nella}, {Rieke}, {Rieke}, {Rix}, {Smith},
  {Sonneborn}, {Stiavelli}, {Stockman}, {Windhorst}, \&
  {Wright}}]{2006SSRv..123..485G}
{Gardner}, J.~P., {Mather}, J.~C., {Clampin}, M., {et~al.} 2006, \ssr, 123,
  485, \dodoi{10.1007/s11214-006-8315-7}

\bibitem[{{Gardner} {et~al.}(2023){Gardner}, {Mather}, {Abbott}, {Abell},
  {Abernathy}, {Abney}, {Abraham}, {Abraham}, {Abul-Huda}, {Acton}, {Adams},
  {Adams}, {Adler}, {Adriaensen}, {Aguilar}, {Ahmed}, {Ahmed}, {Ahmed},
  {Albat}, {Albert}, {Alberts}, {Aldridge}, {Allen}, {Allen}, {Altenburg},
  {Altunc}, {Alvarez}, {{\'A}lvarez-M{\'a}rquez}, {Alves de Oliveira},
  {Ambrose}, {Anandakrishnan}, {Andersen}, {Anderson}, {Anderson}, {Anderson},
  {Anderson}, {Aprea}, {Archer}, {Arenberg}, {Argyriou}, {Arribas}, {Artigau},
  {Arvai}, {Atcheson}, {Atkinson}, {Averbukh}, {Aymergen}, {Bacinski},
  {Baggett}, {Bagnasco}, {Baker}, {Balzano}, {Banks}, {Baran}, {Barker},
  {Barrett}, {Barringer}, {Barto}, {Bast}, {Baudoz}, {Baum}, {Beatty},
  {Beaulieu}, {Bechtold}, {Beck}, {Beddard}, {Beichman}, {Bellagama}, {Bely},
  {Berger}, {Bergeron}, {Bernier}, {Bertch}, {Beskow}, {Betz}, {Biagetti},
  {Birkmann}, {Bjorklund}, {Blackwood}, {Blazek}, {Blossfeld}, {Bluth},
  {Boccaletti}, {Boegner}, {Bohlin}, {Boia}, {B{\"o}ker}, {Bonaventura},
  {Bond}, {Bosley}, {Boucarut}, {Bouchet}, {Bouwman}, {Bower}, {Bowers},
  {Bowers}, {Boyce}, {Boyer}, {Boyer}, {Boyer}, {Boyer}, {Bradley}, {Brady},
  {Brandl}, {Brannen}, {Breda}, {Bremmer}, {Brennan}, {Bresnahan}, {Bright},
  {Broiles}, {Bromenschenkel}, {Brooks}, {Brooks}, {Brown}, {Brown}, {Brown},
  {Bruce}, {Bryson}, {Bujanda}, {Bullock}, {Bunker}, {Bureo}, {Burt}, {Bush},
  {Bushouse}, {Bussman}, {Cabaud}, {Cale}, {Calhoon}, {Calvani}, {Canipe},
  {Caputo}, {Cara}, {Carey}, {Case}, {Cesari}, {Cetorelli}, {Chance},
  {Chandler}, {Chaney}, {Chapman}, {Charlot}, {Chayer}, {Cheezum}, {Chen},
  {Chen}, {Cherinka}, {Chichester}, {Chilton}, {Chittiraibalan}, {Clampin},
  {Clark}, {Clark}, {Clark}, {Claybrooks}, {Cleveland}, {Cohen}, {Cohen},
  {Col{\'o}n}, {Coleman}, {Colina}, {Comber}, {Comeau}, {Comer}, {Conde Reis},
  {Connolly}, {Conroy}, {Contos}, {Contreras}, {Cook}, {Cooper}, {Cooper},
  {Correia}, {Correnti}, {Cossou}, {Costanza}, {Coulais}, {Cox}, {Coyle},
  {Cracraft}, {Crew}, {Curtis}, {Cusveller}, {Da Costa Maciel}, {Dailey},
  {Daugeron}, {Davidson}, {Davies}, {Davis}, {Davis}, {Day}, {de Chambure}, {de
  Jong}, {De Marchi}, {Dean}, {Decker}, {Delisa}, {Dell}, {Dellagatta},
  {Dembinska}, {Demosthenes}, {Dencheva}, {Deneu}, {DePriest}, {Deschenes},
  {Dethienne}, {Detre}, {Diaz}, {Dicken}, {DiFelice}, {Dillman}, {Disharoon},
  {Dixon}, {Doggett}, {Dominguez}, {Donaldson}, {Doria-Warner}, {Santos},
  {Doty}, {Douglas}, {Doyon}, {Dressler}, {Driggers}, {Driggers}, {Dunn},
  {DuPrie}, {Dupuis}, {Durning}, {Dutta}, {Earl}, {Eccleston}, {Ecobichon},
  {Egami}, {Ehrenwinkler}, {Eisenhamer}, {Eisenhower}, {Eisenstein}, {El
  Hamel}, {Elie}, {Elliott}, {Elliott}, {Engesser}, {Espinoza}, {Etienne},
  {Etxaluze}, {Evans}, {Fabreguettes}, {Falcolini}, {Falini}, {Fatig},
  {Feeney}, {Feinberg}, {Fels}, {Ferdous}, {Ferguson}, {Ferrarese}, {Ferreira},
  {Ferruit}, {Ferry}, {Filippazzo}, {Firre}, {Fix}, {Flagey}, {Flanagan},
  {Fleming}, {Florian}, {Flynn}, {Foiadelli}, {Fontaine}, {Fontanella},
  {Forshay}, {Fortner}, {Fox}, {Framarini}, {Francisco}, {Franck}, {Franx},
  {Franz}, {Friedman}, {Friend}, {Frost}, {Fu}, {Fullerton}, {Gaillard},
  {Galkin}, {Gallagher}, {Galyer}, {Garc{\'\i}a Mar{\'\i}n}, {Gardner},
  {Garland}, {Garrett}, {Gasman}, {G{\'a}sp{\'a}r}, {Gastaud}, {Gaudreau},
  {Gauthier}, {Geers}, {Geithner}, {Gennaro}, {Gerber}, {Gereau}, {Giampaoli},
  {Giardino}, {Gibbons}, {Gilbert}, {Gilman}, {Girard}, {Giuliano}, {Gkountis},
  {Glasse}, {Glassmire}, {Glauser}, {Glazer}, {Goldberg}, {Golimowski},
  {Gonzaga}, {Gordon}, {Gordon}, {Goudfrooij}, {Gough}, {Graham}, {Grau},
  {Green}, {Greene}, {Greene}, {Greenfield}, {Greenhouse}, {Greve}, {Greville},
  {Grimaldi}, {Groe}, {Groebner}, {Grumm}, {Grundy}, {G{\"u}del}, {Guillard},
  {Guldalian}, {Gunn}, {Gurule}, {Gutman}, {Guy}, {Guyot}, {Hack}, {Haderlein},
  {Hagan}, {Hagedorn}, {Hainline}, {Haley}, {Hami}, {Hamilton}, {Hammann},
  {Hammel}, {Hanley}, {Hansen}, {Hardy}, {Harnisch}, {Harr}, {Harris}, {Hart},
  {Hartig}, {Hasan}, {Hashim}, {Hashimoto}, {Haskins}, {Hawkins}, {Hayden},
  {Hayden}, {Healy}, {Hecht}, {Heeg}, {Hejal}, {Helm}, {Hengemihle}, {Henning},
  {Henry}, {Henry}, {Henshaw}, {Hernandez}, {Herrington}, {Heske}, {Hesman},
  {Hickey}, {Hilbert}, {Hines}, {Hinz}, {Hirsch}, {Hitcho}, {Hodapp}, {Hodge},
  {Hoffman}, {Holfeltz}, {Holler}, {Hoppa}, {Horner}, {Howard}, {Howard},
  {Huber}, {Hunkeler}, {Hunter}, {Hunter}, {Hurd}, {Hurst}, {Hutchings},
  {Hylan}, {Ignat}, {Illingworth}, {Irish}, {Isaacs}, {Jackson}, {Jaffe},
  {Jahic}, {Jahromi}, {Jakobsen}, {James}, {James}, {James}, {Jamieson},
  {Jandra}, {Jayawardhana}, {Jedrzejewski}, {Jeffers}, {Jensen}, {Joanne},
  {Johns}, {Johnson}, {Johnson}, {Johnson}, {Johnson}, {Johnson}, {Johnson},
  {Johnstone}, {Jollet}, {Jones}, {Jones}, {Jones}, {Jones}, {Jones}, {Jordan},
  {Jordan}, {Jue}, {Jurkowski}, {Justis}, {Justtanont}, {Kaleida}, {Kalirai},
  {Kalmanson}, {Kaltenegger}, {Kammerer}, {Kan}, {Kanarek}, {Kao}, {Karakla},
  {Karl}, {Kassin}, {Kauffman}, {Kavanagh}, {Kelley}, {Kelly}, {Kendrew},
  {Kennedy}, {Kenny}, {Keski-Kuha}, {Keyes}, {Khan}, {Kidwell}, {Kimble},
  {King}, {King}, {Kinzel}, {Kirk}, {Kirkpatrick}, {Klaassen}, {Klingemann},
  {Klintworth}, {Knapp}, {Knight}, {Knollenberg}, {Knutsen}, {Koehler},
  {Koekemoer}, {Kofler}, {Kontson}, {Kovacs}, {Kozhurina-Platais}, {Krause},
  {Kriss}, {Krist}, {Kristoffersen}, {Krogel}, {Krueger}, {Kulp}, {Kumari},
  {Kwan}, {Kyprianou}, {Labador}, {Labiano}, {Lafreni{\`e}re}, {Lagage},
  {Laidler}, {Laine}, {Laird}, {Lajoie}, {Lallo}, {Lam}, {LaMassa}, {Lambros},
  {Lampenfield}, {Lander}, {Langston}, {Larson}, {Larson}, {LaVerghetta},
  {Law}, {Lawrence}, {Lee}, {Lee}, {Lee}, {Leisenring}, {Leveille}, {Levenson},
  {Levi}, {Levine}, {Lewis}, {Lewis}, {Lewis}, {Libralato}, {Lidon},
  {Liebrecht}, {Lightsey}, {Lilly}, {Lim}, {Lim}, {Ling}, {Link}, {Link},
  {Lipinski}, {Liu}, {Lo}, {Lobmeyer}, {Logue}, {Long}, {Long}, {Long}, {Long},
  {L{\'o}pez-Caniego}, {Lotz}, {Love-Pruitt}, {Lubskiy}, {Luers}, {Luetgens},
  {Luevano}, {Lui}, {Lund}, {Lundquist}, {Lunine}, {L{\"u}tzgendorf}, {Lynch},
  {MacDonald}, {MacDonald}, {Macias}, {Macklis}, {Maghami}, {Maharaja},
  {Maiolino}, {Makrygiannis}, {Malla}, {Malumuth}, {Manjavacas}, {Marini},
  {Marrione}, {Marston}, {Martel}, {Martin}, {Martin}, {Martinez}, {Maschmann},
  {Masci}, {Masetti}, {Maszkiewicz}, {Matthews}, {Matuskey}, {McBrayer},
  {McCarthy}, {McCaughrean}, {McClare}, {McClare}, {McCloskey}, {McClurg},
  {McCoy}, {McElwain}, {McGregor}, {McGuffey}, {McKay}, {McKenzie}, {McLean},
  {McMaster}, {McNeil}, {De Meester}, {Mehalick}, {Meixner}, {Mel{\'e}ndez},
  {Menzel}, {Menzel}, {Merz}, {Mesterharm}, {Meyer}, {Meyett}, {Meza},
  {Midwinter}, {Milam}, {Miller}, {Miller}, {Miskey}, {Misselt}, {Mitchell},
  {Mohan}, {Montoya}, {Moran}, {Morishita}, {Moro-Mart{\'\i}n}, {Morrison},
  {Morrison}, {Morse}, {Moschos}, {Moseley}, {Mosier}, {Mosner}, {Mountain},
  {Muckenthaler}, {Mueller}, {Mueller}, {Muhiem}, {M{\"u}hlmann}, {Mullally},
  {Mullen}, {Munger}, {Murphy}, {Murray}, {Muzerolle}, {Mycroft}, {Myers},
  {Myers}, {Myers}, {Myers}, {Myrick}, {Nagle}, {Nayak}, {Naylor}, {Neff},
  {Nelan}, {Nella}, {Nguyen}, {Nguyen}, {Nickson}, {Nidhiry}, {Niedner},
  {Nieto-Santisteban}, {Nikolov}, {Nishisaka}, {Noriega-Crespo}, {Nota},
  {O'Mara}, {Oboryshko}, {O'Brien}, {Ochs}, {Offenberg}, {Ogle}, {Ohl},
  {Olmsted}, {Osborne}, {O'Shaughnessy}, {{\"O}stlin}, {O'Sullivan}, {Otor},
  {Ottens}, {Ouellette}, {Outlaw}, {Owens}, {Pacifici}, {Page}, {Paranilam},
  {Park}, {Parrish}, {Paschal}, {Patapis}, {Patel}, {Patrick}, {Pattishall},
  {Paul}, {Paul}, {Pauly}, {Pavlovsky}, {Pe{\~n}a-Guerrero}, {Pedder}, {Peek},
  {Pelham}, {Penanen}, {Perriello}, {Perrin}, {Perrine}, {Perrygo}, {Peslier},
  {Petach}, {Peterson}, {Pfarr}, {Pierson}, {Pietraszkiewicz}, {Pilchen},
  {Pipher}, {Pirzkal}, {Pitman}, {Player}, {Plesha}, {Plitzke}, {Pohner},
  {Poletis}, {Pollizzi}, {Polster}, {Pontius}, {Pontoppidan}, {Porges},
  {Potter}, {Prescott}, {Proffitt}, {Pueyo}, {Quispe Neira}, {Radich}, {Rager},
  {Rameau}, {Ramey}, {Ramos Alarcon}, {Rampini}, {Rapp}, {Rashford},
  {Rauscher}, {Ravindranath}, {Rawle}, {Rawlings}, {Ray}, {Regan}, {Rehm},
  {Rehm}, {Reid}, {Reis}, {Renk}, {Reoch}, {Ressler}, {Rest}, {Reynolds},
  {Richon}, {Richon}, {Ridgaway}, {Riedel}, {Rieke}, {Rieke}, {Rifelli},
  {Rigby}, {Riggs}, {Ringel}, {Ritchie}, {Rix}, {Robberto}, {Robinson},
  {Robinson}, {Robinson}, {Rock}, {Rodriguez}, {Rodr{\'\i}guez del Pino},
  {Roellig}, {Rohrbach}, {Roman}, {Romelfanger}, {Romo}, {Rosales}, {Rose},
  {Roteliuk}, {Roth}, {Rothwell}, {Rouzaud}, {Rowe}, {Rowlands}, {Roy},
  {Royer}, {Rui}, {Rumler}, {Rumpl}, {Russ}, {Ryan}, {Ryan}, {Saad}, {Sabata},
  {Sabatino}, {Sabbi}, {Sabelhaus}, {Sabia}, {Sahu}, {Saif}, {Salvignol},
  {Samara-Ratna}, {Samuelson}, {Sanders}, {Sappington}, {Sargent}, {Sauer},
  {Savadkin}, {Sawicki}, {Schappell}, {Scheffer}, {Scheithauer}, {Scherer},
  {Schiff}, {Schlawin}, {Schmeitzky}, {Schmitz}, {Schmude}, {Schneider},
  {Schreiber}, {Schroeven-Deceuninck}, {Schultz}, {Schwab}, {Schwartz},
  {Scoccimarro}, {Scott}, {Scott}, {Seaton}, {Seely}, {Seery}, {Seidleck},
  {Sembach}, {Shanahan}, {Shaughnessy}, {Shaw}, {Shay}, {Sheehan}, {Sheth},
  {Shih}, {Shivaei}, {Siegel}, {Sienkiewicz}, {Simmons}, {Simon}, {Sirianni},
  {Sivaramakrishnan}, {Slade}, {Sloan}, {Slocum}, {Slowinski}, {Smith},
  {Smith}, {Smith}, {Smith}, {Smith}, {Smith}, {Smolik}, {Soderblom}, {Sohn},
  {Sokol}, {Sonneborn}, {Sontag}, {Sooy}, {Soummer}, {Southwood}, {Spain},
  {Sparmo}, {Speer}, {Spencer}, {Sprofera}, {Stallcup}, {Stanley},
  {Stansberry}, {Stark}, {Starr}, {Stassi}, {Steck}, {Steeley}, {Stephens},
  {Stephenson}, {Stewart}, {Stiavelli}, {}, {Strada}, {Straughn}, {Streetman},
  {Strickland}, {Strobele}, {Stuhlinger}, {Stys}, {Such}, {Sukhatme},
  {Sullivan}, {Sullivan}, {Sumner}, {Sun}, {Sunnquist}, {Swade}, {Swam},
  {Swenton}, {Swoish}, {Tam Litten}, {Tamas}, {Tao}, {Taylor}, {Taylor}, {te
  Plate}, {Van Tea}, {Teague}, {Telfer}, {Temim}, {Texter}, {Thatte},
  {Thompson}, {Thompson}, {Thomson}, {Thronson}, {Tierney}, {Tikkanen},
  {Tinnin}, {Tippet}, {Todd}, {Tran}, {Trauger}, {Trejo}, {Vinh Truong},
  {Tsukamoto}, {Tufail}, {Tumlinson}, {Tustain}, {Tyra}, {Ubeda}, {Underwood},
  {Uzzo}, {Vaclavik}, {Valenduc}, {Valenti}, {Van Campen}, {van de Wetering},
  {Van Der Marel}, {van Haarlem}, {Vandenbussche}, {van Dishoeck},
  {Vanterpool}, {Vernoy}, {Vila Costas}, {Volk}, {Voorzaat}, {Voyton}, {Vydra},
  {Waddy}, {Waelkens}, {Wahlgren}, {Walker}, {Wander}, {Warfield}, {Warner},
  {Wasiak}, {Wasiak}, {Wehner}, {Weiler}, {Weilert}, {Weiss}, {Wells}, {Welty},
  {Wheate}, {Wheeler}, {White}, {Whitehouse}, {Whiteleather}, {Whitman},
  {Williams}, {Willmer}, {Willott}, {Willoughby}, {Wilson}, {Wilson}, {Wilson},
  {Windhorst}, {Wislowski}, {Wolfe}, {Wolfe}, {Wolff}, {Wondel}, {Woo},
  {Woods}, {Worden}, {Workman}, {Wright}, {Wu}, {Wu}, {Wun}, {Wymer},
  {Yadetie}, {Yan}, {Yang}, {Yates}, {Yeager}, {Yerger}, {Young}, {Young},
  {Yu}, {Yu}, {Zak}, {Zeidler}, {Zepp}, {Zhou}, {Zincke}, {Zonak}, \&
  {Zondag}}]{2023PASP..135f8001G}
{Gardner}, J.~P., {Mather}, J.~C., {Abbott}, R., {et~al.} 2023, \pasp, 135,
  068001, \dodoi{10.1088/1538-3873/acd1b5}

\bibitem[{{Hinkley} {et~al.}(2015){Hinkley}, {Kraus}, {Ireland}, {Cheetham},
  {Carpenter}, {Tuthill}, {Lacour}, {Evans}, \&
  {Haubois}}]{2015ApJ...806L...9H}
{Hinkley}, S., {Kraus}, A.~L., {Ireland}, M.~J., {et~al.} 2015, \apjl, 806, L9,
  \dodoi{10.1088/2041-8205/806/1/L9}

\bibitem[{{Hinkley} {et~al.}(2022){Hinkley}, {Carter}, {Ray}, {Skemer},
  {Biller}, {Choquet}, {Millar-Blanchaer}, {Sallum}, {Miles}, {Whiteford},
  {Patapis}, {Perrin}, {Pueyo}, {Schneider}, {Stapelfeldt}, {Wang},
  {Ward-Duong}, {Bowler}, {Boccaletti}, {Girard}, {Hines}, {Kalas}, {Kammerer},
  {Kervella}, {Leisenring}, {Pantin}, {Zhou}, {Meyer}, {Liu}, {Bonnefoy},
  {Currie}, {McElwain}, {Metchev}, {Wyatt}, {Absil}, {Adams}, {Barman},
  {Baraffe}, {Bonavita}, {Booth}, {Bryan}, {Chauvin}, {Chen}, {Danielski}, {De
  Furio}, {Factor}, {Fitzgerald}, {Fortney}, {Grady}, {Greenbaum}, {Henning},
  {Hoch}, {Janson}, {Kennedy}, {Kenworthy}, {Kraus}, {Kuzuhara}, {Lagage},
  {Lagrange}, {Launhardt}, {Lazzoni}, {Lloyd}, {Marino}, {Marley}, {Martinez},
  {Marois}, {Matthews}, {Matthews}, {Mawet}, {Mazoyer}, {Phillips}, {Petrus},
  {Quanz}, {Quirrenbach}, {Rameau}, {Rebollido}, {Rickman}, {Samland},
  {Sargent}, {Schlieder}, {Sivaramakrishnan}, {Stone}, {Tamura}, {Tremblin},
  {Uyama}, {Vasist}, {Vigan}, {Wagner}, \& {Ygouf}}]{2022PASP..134i5003H}
{Hinkley}, S., {Carter}, A.~L., {Ray}, S., {et~al.} 2022, \pasp, 134, 095003,
  \dodoi{10.1088/1538-3873/ac77bd}

\bibitem[{{Hirata} \& {Choi}(2020)}]{2020PASP..132a4501H}
{Hirata}, C.~M., \& {Choi}, A. 2020, \pasp, 132, 014501,
  \dodoi{10.1088/1538-3873/ab44f7}

\bibitem[{{Ireland}(2013)}]{2013MNRAS.433.1718I}
{Ireland}, M.~J. 2013, \mnras, 433, 1718, \dodoi{10.1093/mnras/stt859}

\bibitem[{{Ireland} \& {Kraus}(2008)}]{2008ApJ...678L..59I}
{Ireland}, M.~J., \& {Kraus}, A.~L. 2008, \apjl, 678, L59,
  \dodoi{10.1086/588216}

\bibitem[{{Jennison}(1958)}]{1958MNRAS.118..276J}
{Jennison}, R.~C. 1958, \mnras, 118, 276, \dodoi{10.1093/mnras/118.3.276}

\bibitem[{{Kammerer} {et~al.}(2019){Kammerer}, {Ireland}, {Martinache}, \&
  {Girard}}]{2019MNRAS.486..639K}
{Kammerer}, J., {Ireland}, M.~J., {Martinache}, F., \& {Girard}, J.~H. 2019,
  \mnras, 486, 639, \dodoi{10.1093/mnras/stz882}

\bibitem[{{Kammerer} {et~al.}(2022){Kammerer}, {Girard}, {Carter}, {Perrin},
  {Cooper}, {Thatte}, {Vandal}, {Leisenring}, {Wang}, {Balmer},
  {Sivaramakrishnan}, {Pueyo}, {Ward-Duong}, {Sunnquist}, \& {Adams
  Redai}}]{2022SPIE12180E..3NK}
{Kammerer}, J., {Girard}, J., {Carter}, A.~L., {et~al.} 2022, in Society of
  Photo-Optical Instrumentation Engineers (SPIE) Conference Series, Vol. 12180,
  Space Telescopes and Instrumentation 2022: Optical, Infrared, and Millimeter
  Wave, ed. L.~E. {Coyle}, S.~{Matsuura}, \& M.~D. {Perrin}, 121803N,
  \dodoi{10.1117/12.2628865}

\bibitem[{{Monnier}(1999)}]{1999PhDT........19M}
{Monnier}, J.~D. 1999, PhD thesis, UNIVERSITY OF CALIFORNIA, BERKELEY

\bibitem[{{Plazas} {et~al.}(2018){Plazas}, {Shapiro}, {Smith}, {Huff}, \&
  {Rhodes}}]{2018PASP..130f5004P}
{Plazas}, A.~A., {Shapiro}, C., {Smith}, R., {Huff}, E., \& {Rhodes}, J. 2018,
  \pasp, 130, 065004, \dodoi{10.1088/1538-3873/aab820}

\bibitem[{Quenouille(1949)}]{10.1214/aoms/1177729989}
Quenouille, M.~H. 1949, The Annals of Mathematical Statistics, 20, 355 ,
  \dodoi{10.1214/aoms/1177729989}

\bibitem[{{Rauscher} {et~al.}(2017){Rauscher}, {Arendt}, {Fixsen},
  {Greenhouse}, {Lander}, {Lindler}, {Loose}, {Moseley}, {Mott}, {Wen},
  {Wilson}, \& {Xenophontos}}]{2017PASP..129j5003R}
{Rauscher}, B.~J., {Arendt}, R.~G., {Fixsen}, D.~J., {et~al.} 2017, \pasp, 129,
  105003, \dodoi{10.1088/1538-3873/aa83fd}

\bibitem[{{Ray} {et~al.}(2023{\natexlab{a}}){Ray}, {Hinkley}, {Sallum},
  {Bonavita}, {Squicciarini}, {Carter}, \& {Lazzoni}}]{2023MNRAS.519.2718R}
{Ray}, S., {Hinkley}, S., {Sallum}, S., {et~al.} 2023{\natexlab{a}}, \mnras,
  519, 2718, \dodoi{10.1093/mnras/stac3425}

\bibitem[{{Ray} {et~al.}(2023{\natexlab{b}}){Ray}, {Sallum}, \&
  {Hinkley}}]{ray_subm}
{Ray}, S., {Sallum}, S., \& {Hinkley}, S. e.~a. 2023{\natexlab{b}}, \apj

\bibitem[{{Sallum} \& {Eisner}(2017)}]{2017ApJS..233....9S}
{Sallum}, S., \& {Eisner}, J. 2017, The Astrophysical Journal Supplement
  Series, 233, 9, \dodoi{10.3847/1538-4365/aa90bb}

\bibitem[{{Sallum} {et~al.}(2023){Sallum}, {Eisner}, {Skemer}, \&
  {Murray-Clay}}]{2023ApJ...953...55S}
{Sallum}, S., {Eisner}, J., {Skemer}, A., \& {Murray-Clay}, R. 2023, \apj, 953,
  55, \dodoi{10.3847/1538-4357/ace16c}

\bibitem[{{Sallum} {et~al.}(2021){Sallum}, {Eisner}, {Stone}, {Dietrich},
  {Hinz}, \& {Spalding}}]{2021AJ....161...28S}
{Sallum}, S., {Eisner}, J.~A., {Stone}, J.~M., {et~al.} 2021, \aj, 161, 28,
  \dodoi{10.3847/1538-3881/abc957}

\bibitem[{{Sallum} {et~al.}(2022){Sallum}, {Ray}, \&
  {Hinkley}}]{2022SPIE12183E..2MS}
{Sallum}, S., {Ray}, S., \& {Hinkley}, S. 2022, in Society of Photo-Optical
  Instrumentation Engineers (SPIE) Conference Series, Vol. 12183, Optical and
  Infrared Interferometry and Imaging VIII, ed. A.~{M{\'e}rand}, S.~{Sallum},
  \& J.~{Sanchez-Bermudez}, 121832M, \dodoi{10.1117/12.2630401}

\bibitem[{{Sallum} \& {Skemer}(2019)}]{2019JATIS...5a8001S}
{Sallum}, S., \& {Skemer}, A. 2019, Journal of Astronomical Telescopes,
  Instruments, and Systems, 5, 018001, \dodoi{10.1117/1.JATIS.5.1.018001}

\bibitem[{{Sallum} {et~al.}(2019){Sallum}, {Skemer}, {Eisner}, {van der Marel},
  {Sheehan}, {Close}, {Ireland}, {Males}, {Morzinski}, {Bailey}, {Briguglio},
  \& {Puglisi}}]{2019ApJ...883..100S}
{Sallum}, S., {Skemer}, A.~J., {Eisner}, J.~A., {et~al.} 2019, \apj, 883, 100,
  \dodoi{10.3847/1538-4357/ab3dae}

\bibitem[{{Schlawin} {et~al.}(2020){Schlawin}, {Leisenring}, {Misselt},
  {Greene}, {McElwain}, {Beatty}, \& {Rieke}}]{2020AJ....160..231S}
{Schlawin}, E., {Leisenring}, J., {Misselt}, K., {et~al.} 2020, \aj, 160, 231,
  \dodoi{10.3847/1538-3881/abb811}

\bibitem[{{Sivaramakrishnan} {et~al.}(2012){Sivaramakrishnan},
  {Lafreni{\`e}re}, {Ford}, {McKernan}, {Cheetham}, {Greenbaum}, {Tuthill},
  {Lloyd}, {Ireland}, {Doyon}, {Beaulieu}, {Martel}, {Koekemoer}, {Martinache},
  \& {Teuben}}]{2012SPIE.8442E..2SS}
{Sivaramakrishnan}, A., {Lafreni{\`e}re}, D., {Ford}, K.~E.~S., {et~al.} 2012,
  in \procspie, Vol. 8442, Space Telescopes and Instrumentation 2012: Optical,
  Infrared, and Millimeter Wave, 84422S, \dodoi{10.1117/12.925565}

\bibitem[{{Sivaramakrishnan} {et~al.}(2023){Sivaramakrishnan}, {Tuthill},
  {Lloyd}, {Greenbaum}, {Thatte}, {Cooper}, {Vandal}, {Kammerer},
  {Sanchez-Bermudez}, {Pope}, {Blakely}, {Albert}, {Cook}, {Johnstone},
  {Martel}, {Volk}, {Soulain}, {Artigau}, {Lafreni{\`e}re}, {Willott},
  {Parmentier}, {Ford}, {McKernan}, {Vila}, {Rowlands}, {Doyon}, {Beaulieu},
  {Desdoigts}, {Fullerton}, {De Furio}, {Goudfrooij}, {Holfeltz}, {LaMassa},
  {Maszkiewicz}, {Meyer}, {Perrin}, {Pueyo}, {Sahlmann}, {Sohn}, {Teixeira}, \&
  {Zheng}}]{2023PASP..135a5003S}
{Sivaramakrishnan}, A., {Tuthill}, P., {Lloyd}, J.~P., {et~al.} 2023, \pasp,
  135, 015003, \dodoi{10.1088/1538-3873/acaebd}

\bibitem[{{Soulain} {et~al.}(2020){Soulain}, {Sivaramakrishnan}, {Tuthill},
  {Thatte}, {Volk}, {Cooper}, {Albert}, {Artigau}, {Cook}, {Doyon},
  {Johnstone}, {Lafreni{\`e}re}, \& {Martel}}]{2020SPIE11446E..11S}
{Soulain}, A., {Sivaramakrishnan}, A., {Tuthill}, P., {et~al.} 2020, in Society
  of Photo-Optical Instrumentation Engineers (SPIE) Conference Series, Vol.
  11446, Optical and Infrared Interferometry and Imaging VII, ed. P.~G.
  {Tuthill}, A.~{M{\'e}rand}, \& S.~{Sallum}, 1144611,
  \dodoi{10.1117/12.2560804}

\bibitem[{{Tuthill} {et~al.}(2001){Tuthill}, {Monnier}, \&
  {Danchi}}]{2001Natur.409.1012T}
{Tuthill}, P.~G., {Monnier}, J.~D., \& {Danchi}, W.~C. 2001, \nat, 409, 1012,
  \dodoi{10.1038/35059014}

\bibitem[{{Tuthill} {et~al.}(2000){Tuthill}, {Monnier}, {Danchi}, {Wishnow}, \&
  {Haniff}}]{2000PASP..112..555T}
{Tuthill}, P.~G., {Monnier}, J.~D., {Danchi}, W.~C., {Wishnow}, E.~H., \&
  {Haniff}, C.~A. 2000, \pasp, 112, 555, \dodoi{10.1086/316550}

\end{thebibliography}
\bibliographystyle{aasjournal}

\end{document}